\documentclass[preprint2]{proto}
\usepackage{natbib}
\bibpunct{(}{)}{;}{a}{,}{,}
\usepackage{times}

\begin{document}

\newcommand{\Msunyr}{M$_{\odot}$yr$^{-1}$}
\newcommand{\Mjup}{M$_{\mathrm {Jup}}$}
\newcommand{\Msun}{M$_{\odot}$}
\newcommand{\Mdot}{\dot{M}}
\newcommand{\kms}{km\,s$^{-1}$}
\newcommand{\neii}{[Ne\,{\sc ii}]}
\newcommand{\neiii}{[Ne\,{\sc iii}]}
\newcommand{\arii}{[Ar\,{\sc ii}]}
\newcommand{\oi}{[O\,{\sc i}]}
\newcommand{\sii}{[S\,{\sc ii}]}
\newcommand{\nii}{[N\,{\sc ii}]}


\title{\textbf{\LARGE The Dispersal of Protoplanetary Disks}}

\author {\textbf{\large Richard Alexander}}
\affil{\small\em University of Leicester}

\author {\textbf{\large Ilaria Pascucci}}
\affil{\small\em University of Arizona}

\author {\textbf{\large Sean Andrews}}
\affil{\small\em Harvard-Smithsonian Center for Astrophysics}

\author {\textbf{\large Philip Armitage}}
\affil{\small\em University of Colorado}

\author {\textbf{\large Lucas Cieza}}
\affil{\small\em Universidad Diego Portales}

\begin{abstract}
\baselineskip = 11pt
\leftskip = 0.65in 
\rightskip = 0.65in
\parindent=1pc
{\small \noindent Protoplanetary disks are the sites of planet formation, and the evolution and eventual dispersal of these disks strongly influences the formation of planetary systems. Disk evolution during the planet-forming epoch is driven by accretion and mass-loss due to winds, and in typical environments photoevaporation by high-energy radiation from the central star is likely to dominate final gas disk dispersal. We present a critical review of current theoretical models, and discuss the observations that are used to test these models and inform our understanding of the underlying physics. We also discuss the role disk dispersal plays in shaping planetary systems, considering its influence on both the process(es) of planet formation and the architectures of planetary systems. We conclude by presenting a schematic picture of protoplanetary disk evolution and dispersal, and discussing prospects for future work. 
\\~\\~\\~}
\end{abstract} 


\section{INTRODUCTION}\label{sec:intro}
\noindent The evolution and eventual dispersal of protoplanetary disks play crucial roles in planet formation. Protoplanetary disks are a natural consequence of star formation, spun up by angular momentum conservation during gravitational collapse. The simple fact that these disks are observed to accrete tells us that they evolve, and observations of disk-less stars show that final gas disk dispersal is very efficient. Disk dispersal therefore sets a strict limit on the time-scale for gas-giant planet formation. Removal of disk gas can also alter the disk's chemical composition, which has important implications for planet formation, and as disk clearing halts planet migration it also influences the initial architectures of planetary systems. In this chapter we review the physics of protoplanetary disk dispersal, and its implications for the formation of planetary systems.


\subsection{Observational Constraints on Disk Dispersal}\label{sec:intro_obs}
\noindent Gas-rich protoplanetary disks were discovered more than 25 years ago \citep[e.g.,][]{sb87}, and are now commonly observed. The chapters by {\em Dutrey et al.}, {\em Espaillat et al.}, {\em Pontoppidan et al.} and {\em Testi et al.} present a comprehensive summary of disk observations \citep[see also][]{pt10,wc11}; here we merely highlight the key observations which motivate and constrain theoretical models of disk dispersal. Note also that, for reasons of length, our discussion focuses on stars of approximately solar mass, which are generally better studied than higher- or lower-mass stars.

Young, solar-like stars are traditionally classified either by the slope of their infrared (IR) spectral energy distribution (SED), or by the strength of emission lines in their spectrum. As circumstellar material absorbs stellar radiation and re-emits it at longer wavelengths, a redder SED is broadly associated with more circumstellar dust. Objects with Class II SEDs are therefore inferred to be stars with disks, while near-stellar Class III SEDs are indicative of young stars which have shed their disks. (Class 0 \& I sources are embedded objects, at an earlier evolutionary stage; \citealt{lada87,andre93}.) The major source of optical emission lines from young stars is accretion: objects with bright emission lines (such as H$\alpha$) are referred to as ``classical T Tauri stars'' (CTTs), while similar stars which lack accretion signatures are designated ``weak-lined T Tauri stars'' (WTTs). CTTs generally have Class II SEDs and WTTs Class III SEDs, and although there is not a perfect correspondence between the different classifications we use these terms interchangably.

The dust (solid) component of the disk represents only a small fraction of the disk mass but dominates the opacity, and dust in the disk absorbs stellar radiation and re-emits it at longer wavelengths. Continuum emission at different wavelengths probes dust at different temperatures, and therefore different radii, in the disk: warm dust in the inner few AU is observed in the near-IR, while mm emission traces cold dust in the outer disk. Observations of gas are more challenging, primarily because the majority of the disk mass is cold molecular hydrogen which emits only through weak quadrupole transitions. Line emission from H$_2$, as well as from CO and other trace species (both atomic and molecular), is detected \citep[see, e.g.,][]{najita_ppv,wc11}, but gas in protoplanetary disks is most readily observed through the signatures of accretion on to the stellar surface. 

This wide variety of observational tracers allows us to build up a broad picture of protoplanetary disk evolution. In the youngest clusters ($\lesssim$\,1Myr) the IR excess fraction for single stars is close to 100\%, but this declines dramatically with age and is typically 10\% or less for ages $\gtrsim$\,5Myr \citep[e.g.,][]{haisch01,mamajek09,kraus12}. A similar decline in disk fraction is seen in accretion signatures \citep{fedele10}, and the mass of cold dust and gas in outer disks is also substantially depleted in older clusters \citep[e.g.,][]{mathews12}. Protoplanetary disk lifetimes are therefore inferred to be a few Myr, with order-of-magnitude scatter.

These observations also allow us to make quantitative measurements of disk properties. Resolved observations of CO emission lines show Keplerian rotation profiles on scales of tens to hundreds of AU \citep[e.g.,][]{simon00}. Disk surface densities typically decline with radius (at least at radii $\gtrsim$\,10AU), with power-law indices ($\Sigma$\,$\propto$\,$R^{-p}$) measured to be $p$\,$\simeq$\,0.5--1 \citep{andrews09}. Stellar accretion rates range from $\dot{M}$\,$\gtrsim$\,$10^{-7}$\Msunyr\ to $\lesssim$\,$10^{-10}$\Msunyr\ \citep[e.g.,][]{muzerolle00}, while disk masses estimated from (sub-)mm continuum emission range from $M_{\mathrm d}$\,$\sim$\,0.1\Msun\ to $\lesssim$\,0.001\Msun\ \citep{aw05}. The accretion time-scales inferred from these measurements (i.e., $t$\,$\sim$\,$M_{\mathrm d}/\dot{M}$) are therefore also $\sim$\,Myr, implying that protoplanetary disks evolve substantially during their lifetimes.

Observations of young disk-less stars show that disk dispersal is extremely efficient. Searches for gas around WTTs yield only upper limits: non-detections of H$_2$ ro-vibrational transitions and other mid-IR gas emission lines imply warm gas surface densities $\Sigma$\,$<$\,1g\,cm$^{-2}$ at $\sim$\,AU radii \citep{pascucci06}, while non-detections of H$_2$ fluorescent electronic transitions suggest $\Sigma$\,$<$\,$10^{-5}$g\,cm$^{-2}$ \citep{ingleby09}. This latter limit is $\sim$\,$10^{-7}$ of the surface densities inferred for accreting CTTs, and suggests that gas disk dispersal is almost total. We also see that the various disk and accretion signatures are very strongly correlated and usually vanish together, implying that clearing occurs nearly simultaneously across the entire radial extent of the disk \citep[e.g.,][]{aw05,cieza08}. 


\begin{figure}[t]
 \centering
 \includegraphics[angle=270,width=\hsize]{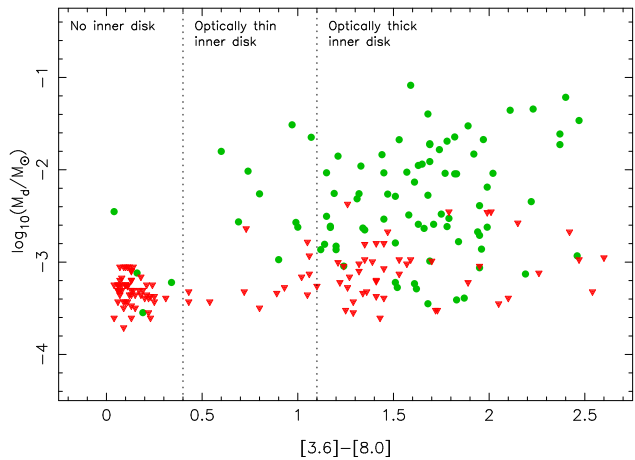}
 \caption{\small Compilation of IR and (sub-)mm data for disks in Taurus-Auriga. The horizontal axis shows the {\it Spitzer} IR color (in magnitudes), while the vertical axis shows the disk mass derived from mm continuum emission (green circles denote detections; red triangles upper limits). IR excesses, which trace inner dust disks, correlate strongly with disk mass, and there is a striking lack of objects with properties between disk-bearing CTTs and disk-less WTTs. [Figure adapted from \citet{cieza08}, using data from \citet{aw05}, \citet{luhman10} and \citet{andrews13}.]}
 \label{fig:IR_mm}
\end{figure}


Finally, relatively few objects show evidence of partial disk clearing (the so-called ``transitional'' disks; see Section \ref{sec:obs_tds}, and the chapter by {\em Espaillat et al.}), and there is a striking dearth of objects with properties intermediate between CTTs and WTTs \citep[e.g.,][]{kh95,duvert00,padgett06}. This suggests that the transition from disk-bearing to disk-less is rapid, as few objects are ``caught in the act'' of disk clearing (see Fig.\ref{fig:IR_mm}). Statistical estimates find the that dispersal time-scale is $\sim$\,10 times shorter than the typical disk lifetime \citep[][]{sp95,ww96,aw05,luhman10,koepferl13}. The mechanism(s) which drive final disk dispersal must therefore efficiently remove both gas and dust, from $<$\,0.1AU to $>$\,100AU, on a time-scale $\sim$\,$10^5$yr, after a disk lifetime of a few Myr.


\subsection{Disk Dispersal Mechanisms}\label{sec:intro_theory}
\noindent Observed disk lifetimes require disks to survive for at least thousands of orbital periods even at large radii. Disks are therefore dynamically long-lived, and evolve relatively slowly. A comprehensive review of protoplanetary disk physics is given by \citet{armitage11}, while \citet{hollenbach_ppiv} discussed disk dispersal mechanisms in detail; here we summarise the key physical processes.

Disk accretion is a major driver of disk evolution, and is generally thought to dominate at early times. The observed evolution of CTT disks on Myr time-scales is broadly consistent with viscous accretion disk theory \citep[e.g.,][]{ss73,lbp74,hartmann98}. In this picture disks evolve due to the exchange (transport) of angular momentum between neighbouring annuli, which is traditionally attributed to viscous stresses. In reality these stresses are due to turbulence (and possibly also laminar magnetic torques) in the disk \citep[e.g.,][see also the chapter by {\em Turner et al.}]{balbus11}, and in the modern interpretation the Shakura-Sunyaev $\alpha$-parameter represents the efficiency of turbulent transport. The characteristic viscous time-scale is
\begin{equation}\label{eq:t_visc}
t_{\nu} = \frac{1}{\alpha \Omega} \left(\frac{H}{R}\right)^{-2} \, ,
\end{equation}
where $\Omega$ is the (Keplerian) orbital frequency, and the disk aspect ratio is typically $H/R$\,$\sim$\,0.1. Observations suggest that $\alpha$\,$\sim$\,$0.01$ \citep[e.g.,][]{hartmann98}, so the local viscous time-scale is at least thousands of orbital periods, and $t_{\nu}$\,$\sim$\,Myr at radii $\gtrsim$\,100AU. Protoplanetary disks are therefore observed to live for at most a few viscous time-scales in their outer regions, implying that some other mechanism drives disk dispersal.

A variety of other mechanisms can remove mass and/or angular momentum from disks. The presence of a binary companion strongly affects disk formation and evolution, particularly for close binaries \citep[e.g.,][]{harris12}. In star clusters disks undergo tidal stripping during close stellar encounters, and can be evaporated or ablated by radiation and winds from massive stars. However, these processes dominate in only a small fraction of disks \citep[e.g.,][]{sc01,adams06}, and the majority of disks in massive clusters have similar properties to those in less hostile environments \citep[e.g.,][]{mann10}. Thus, while environment clearly plays a major role in some cases (see Section \ref{sec:models_ext}), disk evolution around single stars must primarily be due to ``internal'' processes.

Magnetically-launched jets and winds extract both mass and angular momentum, and may drive accretion in protoplanetary disks \citep[e.g.,][]{ks11}. Magnetocentrifugal winds are perhaps the most plausible explanation for protostellar jets and outflows, and may well play a major role in disk evolution at early times.  High-density winds from the star or inner disk may also strip the disk of gas at $\gtrsim$\,AU radii \citep{matsuyama09}.  We defer detailed discussion of magnetic winds to the chapter by {\em Frank et al.}, but consider their role in disk dispersal in Section \ref{sec:models_mhd}.

We also note in passing that the processing of disk material into planets appears not to be a major driver of disk evolution or dispersal. It is now clear that planet formation is ubiquitous, but observations of both the Solar System and exoplanets show that planets account for $\lesssim$\,1\% of the initial disk mass in most systems \citep[e.g.,][]{wright11,mayor13}. Thus, while planet formation may involve a significant fraction of the total mass of heavy elements in the disk, planets represent only a small fraction of the disk mass budget.  In addition, planet formation (by core accretion) is a rather slow process (typically requiring $\sim$\,Myr time-scales; e.g., \citealt{pollack96}), so it seems unlikely that planet formation plays a major role in driving the rapid disk dispersal required by observations.

The final commonly-considered mechanism for disk dispersal is photoevaporation. High-energy radiation (UV and/or X-rays) heats the disk surface to high temperatures ($\sim$$10^3$--$10^4$K), and beyond a few AU this heated layer is unbound and flows from the disk surface as a pressure-driven wind. \citet{hollenbach_ppiv} \& \citet{dullemond_ppv} considered various disk dispersal processes in detail in previous {\em Protostars \& Planets} volumes, and concluded that photoevaporation is the dominant mechanism for removing disk gas at large radii \citep[see also the more recent review by][]{clarke11}. Considerable progress has been made in this field in recent years, and we devote much of this review to the theory and observations of photoevaporative winds. In Section \ref{sec:models} we review the theory of disk dispersal, while Section \ref{sec:obs} discusses the observations which constrain our theoretical models. In Section \ref{sec:conseq} we discuss the implications of these results for the formation and evolution of planetary systems, and we conclude by presenting a schematic picture of the evolution and dispersal of protoplanetary disks.

\bigskip


\section{MODELS OF DISK DISPERSAL}\label{sec:models}
\noindent A variety of processes influence the evolution and dispersal of protoplanetary disks but, for the reasons discussed in Section \ref{sec:intro_theory}, we focus primarily on accretion and photoevaporation. We first review the basic physics of disk photoevaporation, then discuss the current state-of-the-art in theoretical modelling and the observational predictions of these models. We also review recent work suggesting that magnetohydrodynamic (MHD) turbulence can drive winds; mass-loss rates from MHD winds are highly uncertain, but may be high enough to contribute to disk dispersal.


\subsection{Disk photoevaporation: theoretical basics}\label{sec:models_evap}
\noindent The basic principles of disk photoevaporation are readily understood. When high-energy radiation is incident on a disk, its upper layers are heated to well above the midplane temperature. At sufficiently large radius (i.e., high enough in the potential well) the thermal energy of the heated layer exceeds its gravitational binding energy, and the heated gas escapes. The result is a centrifugally-launched, pressure-driven flow, which is referred to as a photoevaporative wind. 

Photoevaporation was applied to disks around young stars as long ago as \citet{bs82}, and the first detailed models were presented by \citet{shu93} and \citet{hollenbach94}. We follow their approach in defining the characteristic length-scale (the ``gravitational radius'') as the (cylindrical) radius where the Keplerian orbital speed is equal to the sound speed of the hot gas:
\begin{equation}
R_{\mathrm g} = \frac{GM_*}{c_{\mathrm s}^2} \, .
\end{equation}
Here $M_*$ is the stellar mass and $c_{\mathrm s}$ is the (isothermal) sound speed of the heated disk surface layer. The simplest case we can consider is an isothermal wind launched from a thin disk in Keplerian rotation, which is analogous to the problem of Compton-heated winds around AGN \citep[e.g.,][]{begelman83}. The flow is launched sub-sonically from the base of the heated atmosphere, and passes through a sonic transition (typically $\sim$\,$R_{\mathrm g}/2$ along each streamline; see Fig.\ref{fig:EUV_wind_structure}) before becoming supersonic at large radii.

A purely pressure-driven wind exerts no torque, and depletes the disk without altering its specific angular momentum. Photoevaporation therefore has two distinct time-scales: the flow time-scale (which by definition is approximately the dynamical time-scale), and the much longer mass-loss time-scale. If the wind structure does not vary rapidly, these time-scales can be considered independently. We can therefore construct dynamical models of photoevaporation in order to determine the mass-loss profile $\dot{\Sigma}(R,t)$, which can then be incorporated into secular disk evolution models as a sink term. 

In the case of an isothermal wind, simple arguments show that the rate of mass-loss per unit area $\dot{\Sigma}(R)$ peaks at $R_{\mathrm c}$\,$\simeq$\,$R_{\mathrm g}/5$ \citep{liffman03,font04,dullemond_ppv}, and this is now commonly referred to in the literature as the ``critical radius'' \citep[see also][]{adams04}. However, even in the isothermal case the flow solution is analytically intractable \citep[as pointed out by][]{begelman83}, as the bulk properties of the flow depend on the local divergence of the streamlines (which is not known {\it a priori}). We can compute solutions numerically if the base density profile $\rho_0(R)$ is known, but in general this is not the case. The base density profile is determined by the balance between radiative heating and cooling, and the heating rate in turn depends on how radiation is transported through the disk atmosphere. The irradiation can be either ``external'' (i.e., from nearby massive stars) or ``internal'' (i.e., from the central star): the former is dominant in the central regions of massive star clusters (see Section \ref{sec:models_ext}), but for the reasons discussed in Section \ref{sec:intro_theory} we focus on the latter. Central star-driven photoevaporation has two limiting cases: i) an optically thick disk, where the atmosphere is irradiated obliquely; ii) a disk with an optically thin inner hole (expected during disk dispersal at late times), where the base of the flow is heated normally (i.e., face-on). Disk photoevaporation is a coupled problem in radiative transfer, thermodynamics and hydrodynamics, but unfortunately full radiation hydrodynamic simulations remain prohibitively expensive. Instead we must use simplified models, with the choice of simplifications depending primarily on the dominant heating mechanism. Three wavelength regimes are particularly relevant to protoplanetary disks: ionizing, extreme-UV radiation (EUV; 13.6--100eV); far-ultraviolet radiation (FUV; 6--13.6eV), which is capable of dissociating H$_2$ and other molecules; and X-rays (0.1--10keV). In the following sections we discuss models of these processes in turn, highlighting the main results and addressing the shortcomings of the different approaches.


\subsection{Models of photoevaporative winds}\label{sec:models_winds}
\subsubsection{EUV heating}
\noindent The simplest of these three cases is EUV heating, where the incident photons are sufficiently energetic to ionize hydrogen atoms. The absorption cross section at the Lyman limit ($h\nu$\,=\,13.6eV; $\lambda$\,=\,912\AA) is very large ($\sigma$\,=\,$6.3\times10^{-18}$cm$^2$) and decreases approximately as $\nu^{-3}$ \citep{osterbrock_agn2}, so the dominant contribution to the ionization rate comes from photons at or close to the threshold energy. The EUV heating rate is therefore not very sensitive to the incident spectrum, and to a good approximation depends only on the ionizing photon luminosity $\Phi$. The resulting radiative transfer problem is analogous to an ionization-bounded H\,{\sc ii} region, and results in a near-isothermal ionized atmosphere (with $T$\,$\simeq$\,$10^4$\,K and $c_{\mathrm s}$\,$\simeq$\,$10$km\,s$^{-1}$), separated from the neutral underlying disk by an ionization front \citep[see discussion in][]{clarke11}. The typical critical radius is therefore
\begin{equation}
R_{\mathrm {c,EUV}} \simeq 1.8 \left(\frac{M_*}{1\mathrm M_{\odot}}\right) \, \mathrm {AU} \, .
\end{equation}
Close to the star (i.e., $R$\,$\ll$\,$R_{\mathrm c}$) the ionized disk atmosphere is bound, but at larger radii the ionization front represents the launching surface for the photoevaporative wind. The radiative transfer problem therefore reduces to one of ionization balance, as specifying the position and density of the ionization front uniquely determines the flow solution.

The first quantitative models of this process (for the case of an optically thick disk) were performed by \citet[][hereafter HJLS94]{hollenbach94}, using ``1+1D'' numerical radiative transfer calculations. Approximately 1/3 of radiative recombinations of hydrogen produce another ionizing photon, and HJLS94 found that this diffuse (recombination) field dominates at the ionization front for all radii of interest. They divided the atmosphere into ``static'' and ``flow'' regions, and showed that the base density profile, $\rho_0(R)$, can be estimated from a Str\"omgren-like condition \citep[see also][]{alexander08a,clarke11}. The heating process is recombination-limited, and consequently the base density scales as $\Phi^{1/2}$. HJLS94's models involved no hydrodynamics, and instead simply estimated the photoevaporation rate by assuming $\dot{\Sigma}(R) = 2 \rho_0(R) c_{\mathrm s}$ beyond $R_{\mathrm g}$.

\begin{figure}[t]
 \centering
 \includegraphics[angle=270,width=\hsize]{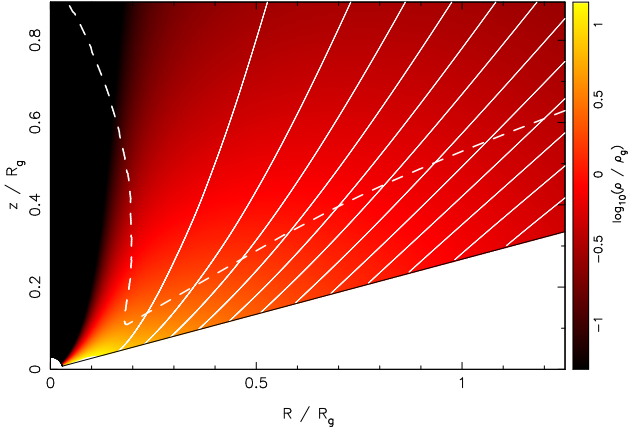}
 \caption{\small Typical structure of a photoevaporative wind.  The color-scale shows the gas density (normalised to the base density at $R$\,$=$\,$R_{\mathrm g}$), with streamlines plotted at 5\% intervals of the total mass flux.  The dashed line shows the location of the sonic surface.  This example shows the structure of an isothermal, EUV-driven wind, from the models used in \citet{pascucci11}.  For a 1\Msun\ star with ionizing luminosity $\Phi$\,$=$\,$10^{41}$photon\,s$^{-1}$, the units correspond to $R_{\mathrm g}$\,$=$\,8.9AU and $n_{\mathrm g}$\,$=$\,$\rho_{\mathrm g}/\mu m_{\mathrm H}$\,$=$\,$2.8\times10^4$cm$^{-3}$.}
 \label{fig:EUV_wind_structure}
\end{figure}

\begin{figure}[t]
 \centering
 \includegraphics[angle=270,width=\hsize]{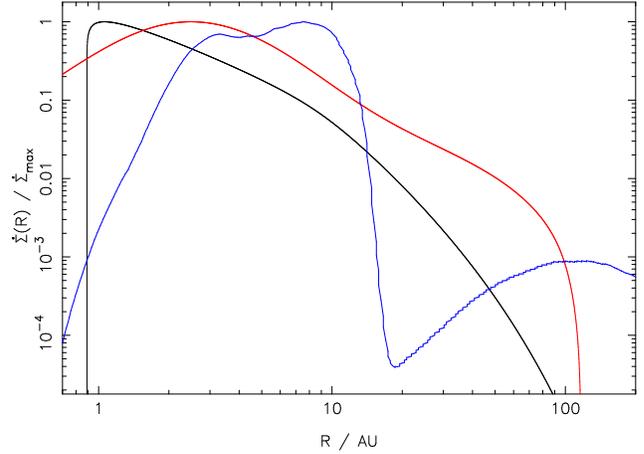}
 \caption{\small Normalised mass-loss profiles for different photoevaporation models around a 1\Msun\ star: EUV- \citep[black line;][]{font04}; X-ray- \citep[red line;][]{owen12}; and FUV-dominated \citep[blue line;][]{gorti09}. Note that for fiducial parameters the absolute mass-loss rates differ by 1--2 orders of magnitude.}
 \label{fig:mdot_profiles}
\end{figure}

\citet{ry97} subsequently introduced numerical hydrodynamics, and studied the effects of dust opacity, in the massive star regime. The first hydrodynamic models of central-star-driven photoevaporation around solar-mass stars were presented by \citet{font04}. These models used 2-D numerical hydrodynamics, assuming an isothermal equation of state and adopting the base density profile $\rho_0(R)$ derived by HJLS94 as a (input) boundary condition. This allows numerical solution of a steady-state wind profile, which in turn quantifies several of the estimates and assumptions made above; the wind structure resulting from such a calculation is shown in Fig.\ref{fig:EUV_wind_structure}. The launch velocity at the base of the flow is typically 0.3--0.4$c_{\mathrm s}$, and the mass-loss profile $\dot{\Sigma}(R)$ peaks at $\simeq$\,0.14$R_{\mathrm g}$ (see Fig.\ref{fig:mdot_profiles}, black line). The total mass-loss per logarithmic interval in radius, $2 \pi R^2\dot{\Sigma}$, peaks at approximately 9AU.  The integrated mass-loss rate over the entire disk is
\begin{eqnarray}
\label{eq:mdot_EUV}
\dot{M}_{\mathrm {w,EUV}} \simeq 1.6 \times 10^{-10} \left(\frac{\Phi}{10^{41}\mathrm s^{-1}}\right)^{1/2} 
\nonumber \\
\times \left(\frac{M_*}{1 \mathrm M_{\odot}}\right)^{1/2} \mathrm M_{\odot}\,\mathrm {yr}^{-1} \, .
\end{eqnarray}
The behaviour of the EUV wind changes significantly in the case of a disk with an optically thin inner hole, as the heating is dominated by direct irradiation from the central star. \citet{alexander06a} studied this problem using both analytic arguments and 2-D numerical hydrodynamics. As the direct field dominates they were able to compute the location of the ionization front ``on-the-fly'' in their hydrodynamic calculations, and find a self-consistent flow solution. In this case \citet{alexander06a} found that
\begin{equation}
\dot{M}_{\mathrm {w}} \simeq 1.4 \times 10^{-9} \left(\frac{\Phi}{10^{41}\mathrm s^{-1}}\right)^{1/2} \left(\frac{R_{\mathrm {in}}}{3 \mathrm {AU}}\right)^{1/2} \mathrm M_{\odot}\,\mathrm {yr}^{-1} \, ,
\end{equation}
where $R_{\mathrm {in}}$ is the radius of the inner disk edge. Direct irradiation therefore increases the wind rate by approximately an order of magnitude, and the mass-loss rate increases for larger inner holes.


From a theoretical perspective, EUV photoevaporation is now well understood, and there is good agreement between analytic and numerical models. The remaining weakness in this approach is that (in the optically thick case) the hydrodynamic models still rely on the radiative transfer solution of HJLS94, and are not strictly self-consistent. This is only a minor issue, however, and is dwarfed by the much larger uncertainty in the input parameters: the ionizing luminosities of TTs are very poorly constrained by observations (see Section \ref{sec:rad_fields}). Moreover, both the accretion columns and any jets or winds close to the star are extremely optically thick to ionizing photons \citep{alexander04a,gorti09,owen12}, so estimating the ionizing flux which reaches the disk at $\sim$AU radii is by far the dominant uncertainty in calculations of EUV photoevaporation.


\subsubsection{X-ray heating}\label{sec:models_xrays}
\noindent We next consider photoevaporation by stellar X-rays. TTs are known to be bright (though highly variable) X-ray sources, with median luminosity $L_{\mathrm X}$$\,\sim$\,$10^{30}$erg\,s$^{-1}$ and a spectrum that typically peaks around 1keV \citep[e.g.,][]{feigelson07}. X-rays have long been known to play an important role in sustaining the level of disk ionization required to drive MHD turbulence \citep{glassgold97,glassgold00}, but recently attention has also turned to the thermal effects of X-ray irradiation. $\sim$keV X-rays are absorbed by K-shell ionization of heavy elements (primarily O, but also C \& Fe), and the resulting photoelectrons then collisionally ionize and/or heat the hydrogen atoms/molecules in the gas. In general the gas and dust temperatures are decoupled, and the dominant cooling channels are i) metal line emission; ii) gas cooling via collisions with grains \citep[e.g.,][]{ercolano08}. For solar abundances the absorption cross section is $\sigma$\,$\simeq$\,$2\times10^{-22}(h\nu/1\mathrm{keV})^{-2.485}$cm$^2$ \citep{glassgold97}.  0.3--1keV X-rays therefore provide significant heating up to a (neutral hydrogen) column depth $N_{\mathrm H}$\,$\sim$\,$10^{21}$--$10^{22}$cm$^{-2}$.

The first models of X-ray photoevaporation were similar in spirit to HJLS94, and used only hydrostatic calculations. \citet{alexander04b} used a simple heating model, and \citet{ercolano08,ercolano09} improved on this approach by using 2-D Monte Carlo radiative transfer \citep[see also][]{gh09}. The resulting disk structure consists of a very tenuous hot ($\sim$\,$ 10^6$K) corona, above a partially ionized atmosphere at $T$\,$\sim$\,$10^3$--$10^4$K. There is a smooth transition from this atmosphere to the (cold) underlying disk, unlike in the EUV case, and the varying temperature means that the flow is not well characterised by a single critical radius. The vertical density gradient is steep, so a small uncertainty in the vertical location of the launch point leads to a large uncertainty in the mass flux. Estimating photoevaporation rates from static calculations is thus fraught with difficulty, and despite deriving similar structures the studies of \citet{alexander04b}, \citet{ercolano08,ercolano09} and \citet{gh09} came to qualitatively different conclusions about the importance of X-ray heating.

This problem was essentially solved by \citet{owen10}, who coupled the radiative transfer models of \citet{ercolano09} to numerical hydrodynamics. These authors used radiative transfer calculations to establish a monotonic relationship between the X-ray ionization parameter ($\xi=L_{\mathrm X}/nd^2$) and the gas temperature, which can be used in lieu of an energy equation in hydrodynamic calculations. This allows the wind structure to be computed numerically, and the steady-state solution can be verified {\em a posteriori} against the radiative transfer code. The photoevaporative flow is launched from the atomic layer, at temperatures $\simeq$3000--5000K, and the flow structure is largely determined by the temperature and density at the sonic point \citep{owen12}.  The resulting mass-loss profile $\dot{\Sigma}(R)$ is broader than in the EUV case, and peaks at $\simeq$3AU (see Fig.\ref{fig:mdot_profiles}, red line); the mass-loss rate $2\pi R^2\dot{\Sigma}$ peaks at 40--60AU. Heating at the base of the flow is dominated by X-rays with $h\nu$\,$\gtrsim$\,0.3--0.4keV, and the integrated wind rate scales almost linearly with $L_{\mathrm X}$ (and is largely insensitive to stellar mass). Assuming a fixed input spectrum with variable luminosity, \cite{owen11,owen12} fit the following scaling relation:
\begin{eqnarray}\label{eq:mdot_X}
\dot{M}_{\mathrm {w,X}} \simeq 6.3 \times 10^{-9} \left(\frac{L_{\mathrm X}}{10^{30}\mathrm {erg\,s}^{-1}}\right)^{1.14}
\nonumber \\
\times \left(\frac{M_*}{1 \mathrm M_{\odot}}\right)^{-0.068} \mathrm M_{\odot}\,\mathrm {yr}^{-1} \, .
\end{eqnarray}
Strikingly, this is $\sim$\,40 times larger than the fiducial EUV-driven wind rate (Equation \ref{eq:mdot_EUV}). As X-ray heating is always local to the initial absorption, the integrated wind rate increases only modestly (by a factor of $\sim$\,2) in the presence of an optically thin disk inner hole. However, \citet{owen12,owen13b} find that X-ray photoevaporation of an inner hole has a dramatic effect if the disk surface density is sufficiently small (typically $\lesssim$\,0.1--1g\,cm$^{-2}$). In this case the physical depth of the X-ray-heated column is comparable to the disk scale-height, rendering the inner edge of the disk dynamically unstable. This instability leads to very rapid (dynamical) dispersal of the disk, and \citet{owen12,owen13b} dub this process ``thermal sweeping''.

The use of the ionization parameter--temperature relation restricts this method to the regime where X-ray heating is dominant, but the resulting wind solutions are robust. As in the EUV case, however, the input radiation field remains a significant source of uncertainty. Although TTs are readily observed in X-rays, observations lack the spectral resolution to be used as inputs for these models. Instead, \citet{ercolano08,ercolano09} and \citet{owen10} created synthetic input spectra, using a coronal emission measure distribution derived primarily from studies of RS CVn-type binaries. It is notable, however, that \citet{gh09} found very different results using a somewhat harder X-ray spectrum.  Some discrepancies between competing radiative transfer codes also remain unresolved, particularly with regard to the temperatures in the X-ray-heated region: the models of Gorti \& Hollenbach consistently predict lower temperatures this region than the models of Ercolano, Owen and collaborators \citep[see, e.g., the discussion in][]{ercolano09}. In addition, as X-rays are primarily absorbed by heavy elements the heating rates are sensitive to assumptions about disk chemistry, and effects such as dust settling have not yet been considered in detail. X-ray photoevaporation models are thus subject to potentially significant systematic uncertainties, and although the individual calculations are mature and self-consistent, further exploration of these issues is desirable.


\subsubsection{FUV heating}\label{sec:models_fuv}
\noindent Finally we consider photoevaporation by non-ionizing, H$_2$-dissociating, FUV radiation. This is by far the most complex case, and is still not fully understood \citep[see][for a more detailed review]{clarke11}. The radiative transfer problem is similar to that in photodissociation regions \citep[PDRs; e.g.,][]{th85}, but is complicated substantially by the disk geometry. The heating/cooling balance in PDRs depends strongly on the ratio of the incident flux to the gas density $G_0/n$. For FUV photoevaporation we are generally in the high $G_0/n$ limit, so dust attenuation of the incident flux dominates and the heated column has a roughly constant depth $N_{\mathrm H}$\,$\sim$\,$10^{21}$--$10^{23}$cm$^{-2}$ that depends primarily on the dust properties (and is comparable to the depth of the X-ray-heated region). Most of the FUV flux is absorbed by dust grains and re-radiated as (IR) continuum, though absorption and re-emission by polycyclic aromatic hydrocarbons (PAHs) is also significant. Gas heating is due to collisions with photo-electrons from grains and PAHs, or FUV-pumping of H$_2$ (followed by fluorescence and collisional de-excitation). The gas and dust temperatures are decoupled at low column density, with gas temperatures ranging from a few hundred to several thousand K \citep[e.g.,][]{adams04}.

FUV heating usually dominates in the case of photoevaporation by nearby massive stars \citep[e.g.,][see also Section \ref{sec:models_ext}]{johnstone98}, but understanding FUV irradiation by the central star remain a work in progress. \citet{gh04,gh08,gh09} have constructed a series of models of this process, using ``1+1-D'' radiative transfer and a thermo-chemical network to compute the structure of the disk atmosphere. The computational expense of these calculations precludes the addition of numerical hydrodynamics; mass fluxes are instead estimated analytically, using a method similar to that of \citet{adams04}. These models use an input spectrum which spans the EUV, FUV \& X-rays, and includes contributions from both coronal emission and the stellar accretion shock (as well as attenuation near the star). The wind is again mainly launched from the atomic layer, but the temperature in the launching region varies substantially, ranging from $>$\,1000K at a few AU to $<$\,100K at $\sim$\,100AU. The resulting mass-loss profile has a peak at 5--10AU, and the FUV irradiation also drives significant mass-loss at large radii ($\gtrsim$\,100AU; see Fig.\ref{fig:mdot_profiles}, blue line). The integrated wind rate depends primarily on the total FUV flux, and for fiducial parameters ($L_{\mathrm {FUV}}$\,$=$\,$5\times10^{31}$erg\,s$^{-1}$, $M_*$\,$=$\,1\Msun) is $3\times 10^{-8}$\Msunyr\ \citep{gh09}. This is again two orders of magnitude larger than the fiducial EUV wind rate, and comparable to the X-ray-dominated wind rates found by \citet{owen10,owen12}.  As much of the mass-loss originates at large radii, FUV photoevaporation can  play a major role in depleting the disk's mass reservoir, and can also potentially truncate disks at radii  $\gtrsim$\,100AU.

However, due to the complexity of this problem these models still suffer from significant uncertainties. In particular, the thermal physics can be very sensitive to the abundance of PAHs in the disk atmosphere. The abundance and depletion of PAHs is highly uncertain \citep[e.g.,][]{geers09}, and recent calculations suggest that changes to the PAH and dust abundances may alter the heating/cooling rates significantly ({\em Gorti et al.}, in prep.). In addition, these models still lack detailed hydrodynamics. Estimating mass fluxes from hydrostatic calculations is problematic (as discussed in Section \ref{sec:models_xrays}), and introduces an additional uncertainty in the derived wind rates. FUV-dominated photoevaporation rates are therefore still uncertain at the order-of-magnitude level, and further work is needed to understand this complex process fully.


\subsubsection{External irradiation}\label{sec:models_ext}
\noindent Our discussion has focused on photoevaporation of disks by their central stars, but it has long been recognised that external irradiation dominates in some cases. The geometry of the radiative transfer problem (essentially plane-parallel irradiation) is much simpler than in the central-star case, and consequently the flow structure is amenable to semi-analytic solution. \citet[][see also \citealt{ry00}]{johnstone98} constructed detailed models of disk evaporation by radiation from nearby O-type stars, as expected in the cores of massive clusters, and found that the heating is dominated by the photospheric EUV and FUV flux. Again the basic picture is that the wind is launched from a PDR on the disk surface, with the thickness of the PDR determined by the incident flux (and hence the distance $d$ from the ionizing source).  For disks in the immediate vicinity of an O-star ($d$\,$\lesssim$\,0.03pc) the ionization front is roughly coincident with the disk surface, and the mass-loss rate is determined by the incident EUV flux. At larger distances, where the ionizing flux is weaker, the PDR thickens and the wind is launched from the neutral, FUV-heated layer. The wind is optically thick to EUV photons, so the ionization front is offset from the base of the flow. For spatially extended disks (with size $R_{\mathrm d}$\,$\gtrsim$\,$R_{\mathrm c}$) the wind is launched almost vertically from the irradiated disk surface, while for more compact (``sub-critical'') disks the wind is launched radially from the disk outer edge \citep{adams04}. The flow subsequently interacts with ionizing photons from the irradiating star, leading to an ionization front with a characteristic cometary shape. Typical mass-loss rates are $\dot{M}_{\mathrm w}$\,$\sim$\,$10^{-7}$\Msunyr, with a strong dependence on the disk size $R_{\mathrm d}$.

External photoevaporation is best seen in the Orion Nebula Cluster (ONC), where a small number ($\sim$\,100) of so-called proplyds (``PROtoPLanetarY DiskS'') are observed in silhouette against the background nebula. These objects show bright emission lines, offset ionization fronts and cometary shapes, and when discovered they were quickly recognised as disks undergoing external photoevaporation \citep[e.g.,][]{odell93,mod96}. The study of proplyds is now relatively mature, and there is excellent agreement between models and observations of their photoevaporative flows \citep[e.g.,][]{sh99,hod99,mesa-delgado12}. The principal factor controlling the long-term evolution of disks subject to external photoevaporation is the initial disk mass \citep{clarke07}, but most disks around solar-mass stars do not experience such harsh environments. Dynamical models of clusters find that external photoevaporation, and other ``environmental'' factors such as stellar encounters, play a significant role in the evolution of only a small fraction of disks \citep[$\lesssim$\,10\%;][]{sc01,adams06}. Observations of disk masses in the ONC suggest that the disks within $d$\,$\lesssim$\,0.3pc of the cluster core are significantly depleted compared to those at larger distances \citep{eisner06,mann10}, consistent with this scenario. Thus, although external photoevaporation drives the evolution of proplyds in the centres of massive clusters, it is not thought to play a major role in the evolution and dispersal of the majority of disks.


\subsection{Coupling to models of disk evolution}{\label{sec:models_evo}
\noindent The mass-loss rates discussed above exceed the observed (stellar) accretion rates of many CTTs (typically $\dot{M}$\,$\sim$\,$10^{-9}$ --$10^{-8}$\Msunyr; e.g., \citealt{hartmann98,muzerolle00}), which suggests that photoevaporation plays a major role in the evolution and dispersal of protoplanetary disks. However, the local mass-loss time-scale ($t$\,$\sim$\,$\Sigma/\dot{\Sigma}$) exceeds the viscous time-scale in much of the disk, so understanding how photoevaporation influences disk evolution requires us to consider these competing processes simultaneously. \citet{clarke01} presented the first such models, combining a one-dimensional viscous accretion disk model with the photoevaporation prescription of HJLS94. In this scenario the low (EUV) wind rate is initially negligible, but the disk accretion rate declines with time and eventually becomes comparable to the mass-loss rate due to photoevaporation. After this time the outer disk is no longer able to re-supply the inner disk, as the mass-loss is concentrated at a particular radius ($\simeq$\,$R_{\mathrm c}$; see Fig.\ref{fig:mdot_profiles}), and all of the accreting material is ``lost'' to the wind. The wind first opens a gap in the disk at $\simeq$\,$R_{\mathrm c}$, and the interior gas then accretes on to the star on its (short) viscous time-scale, removing the inner disk in $\sim$\,$10^5$yr (see Fig.\ref{fig:mdot_time}, black solid line). \citet{alexander06a,alexander06b} subsequently extended this model to consider direct photoevaporation of the outer disk, and found that photoevaporation efficiently clears the entire disk from the inside-out on a time-scale of a few $\times$\,$10^5$yr (after a disk lifetime of a few Myr). This is consistent with the two-time-scale behaviour described in Section \ref{sec:intro_obs}. Moreover, \citet{aa07} found that dust grains are accreted from the inner disk even more rapidly than the gas, confirming that this process efficiently clears both the gas and dust disks. Note, however, that in order for disk clearing to operate in this manner we require that photoevaporation be powered by something {\it other} than the accretion luminosity, as otherwise the wind fails to overcome the accretion flow \citep[e.g.,][]{matsuyama03a}. The source of high-energy photons is usually assumed to be the stellar chromosphere or corona and, due to the role of photoevaporation in precipitating this rapid disk clearing, models of this type are generally referred to as ``UV-switch'' models \citep[after][]{clarke01}.


\begin{figure}[t]
 \centering
 \includegraphics[angle=270,width=\hsize]{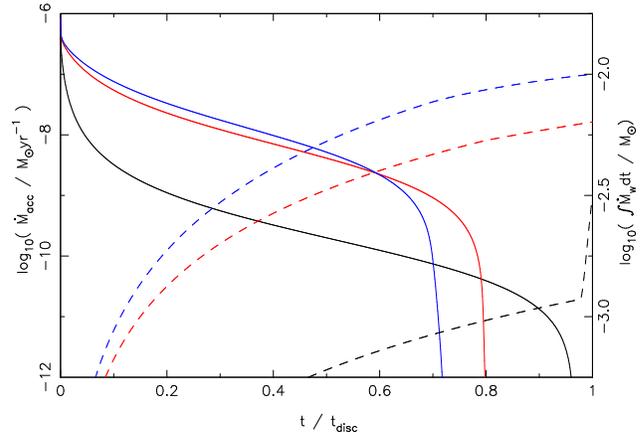}
 \caption{\small Stellar accretion rate (solid lines, left axis) and total mass lost to photoevaporation (dashed lines, right axis) as a function of time for a viscous disk model with different photoevaporation models. The disk model is the median disk from \citet{aa09}, which has $\alpha=0.01$ and an initial disk mass of $10^{-1.5}$\Msun. The various curves show how this disk model evolves when subject to EUV (black lines), X-ray (red) or FUV photoevaporation (blue), using the fiducial wind profiles described in Section \ref{sec:models_winds}. In all cases the accretion rate drops rapidly to zero once photoevaporation triggers inner disk clearing. For clarity, the horizontal axis shows relative time, as the absolute time-scales are primarily a result of the adopted initial conditions and viscosity law. Note, however, that for the same disk parameters the absolute lifetimes for these models differ by factors of several.}
 \label{fig:mdot_time}
\end{figure}


These models considered only EUV photoevaporation, but the evolution is qualitatively similar regardless of the photoevaporation mechanism: once the disk accretion rate drops below the wind rate, photoevaporation takes over and rapidly clears the disk from inside-out \citep[e.g.,][]{gorti09,owen10}. However, the much higher photoevaporation rates predicted by models of X-ray and FUV photoevaporation qualitatively change how the disk evolves. Mass-loss due to EUV photoevaporation is much lower than typical disk accretion rates, and therefore only influences the disk at late times, after it has undergone substantial viscous evolution. By contrast, the X-ray/FUV models of \citet{owen10,owen12} and \citet{gorti09} predict photoevaporation rates that are comparable to the median accretion rate for CTTs ($\sim$$10^{-9}$--$10^{-8}$\Msunyr). In these models photoevaporation therefore represents a significant mass sink even in the early stages of disk evolution ($t$\,$\lesssim$\,1Myr), and the total mass lost to photoevaporation on Myr time-scales may exceed that accreted on to the star (see Fig.\ref{fig:mdot_time}). Such high wind rates challenge the conventional paradigm of viscous accretion in protoplanetary disks, and are arguably the most controversial recent development in the theory of disk evolution and dispersal.

However, despite these important differences between competing models, the major uncertainty in our theory of disk evolution remains our ignorance of how angular momentum is transported. All of the models discussed above adopt simplified viscosity laws, typically by assuming a constant $\alpha$. A constant $\alpha$-parameter is arguably valid as a global average in space and time \citep[e.g.,][]{hartmann98,king07}, but is clearly not an accurate description of the disk microphysics. To date, the only time-dependent models which have attempted to incorporate both photoevaporation and a more physical treatment of angular momentum transport are those of \citet{morishima12} and \citet{bae13}, which considered the evolution of layered disks \citep[e.g.,][]{gammie96,armitage01} subject to X-ray photoevaporation. In this scenario the presence of a dead zone at the disk midplane acts as a bottleneck for the accretion flow, preventing steady accretion through the inner disk at $\gtrsim$\,$10^{-8}$\Msunyr. If the photoevaporation rate exceeds the bottleneck accretion rate, then the wind can open a gap in the disk while a substantial dead zone is still present. The gap generally opens beyond the dead zone, and consequently the inner disk drains much more slowly than in the canonical picture of \citet{clarke01}; \citet{morishima12} argues that this may be the origin of the (small) population of transitional disks known to have large inner dust holes and on-going gas accretion. Real disks may therefore behave rather differently to the simple models described above, but a better understanding of angular momentum transport is required to make significant further progress.

The disk evolution models described here also make a number of other simplifications. In particular, the model initial conditions are highly idealised, and the stellar irradiation which drives photoevaporation is usually assumed to be constant over the disk lifetime. The latter assumption seems plausible (see Section \ref{sec:rad_fields}), but a more realistic treatment of the early stages of disk evolution, incorporating infall and gravitational instabilities, is clearly desirable. If photoevaporation only influences the disk significantly at late times then disk dispersal is largely insensitive to the choice of initial conditions, but this is not necessarily true if the mass-loss rate is high. Moreover, if we are to use these models to further our understanding of planet formation and migration \citep[e.g.,][]{aa09}, then care must be taken to ensure that our treatment of disk formation and early evolution does not influence our results unduly. 


\subsection{Disk photoevaporation: observational predictions}\label{sec:models_pred}
\noindent These models are now relatively mature, and make a series of explicit observational predictions. Many of these predictions are testable with current facilities, offering the opportunity both to identify the mechanisms driving disk dispersal, and to discriminate between competing models. We discuss the theory behind the most important predictions here, and then review the observational evidence in detail in Section \ref{sec:obs}.

As with the models, predictions from disk dispersal models generally take one of two forms: direct diagnostics of the photoevaporative wind, and predictions for the properties of evolving disks (which are usually statistical in nature). The primary diagnostic of disk photoevaporation is line emission from the wind, as the low gas density ($\lesssim$\,$10^6$\,cm$^{-3}$) gives rise to numerous forbidden emission lines. The first detailed calculations of line emission from (EUV) photoevaporative winds driven by the central star were performed by \citet{font04}, who used numerical hydrodynamics and a simplified radiative transfer scheme to compute synthetic line profiles for various optical forbidden lines ([N\,{\sc ii}] 6583\AA, [S\,{\sc ii}] 6716/6731\AA\ and [O\,{\sc i}] 6300\AA). They found that the high flow velocities result in broad line-widths ($\sim$\,30km\,s$^{-1}$) from the ionized species, and showed that for low disk inclinations these lines are blue-shifted by 5--10km\,s$^{-1}$. Their predicted line fluxes ($\sim$\,$10^{-5}$L$_{\odot}$) were consistent with observations of TTs \citep[e.g.,][]{heg95}, but they noted that an EUV wind cannot be the origin of the observed [O\,{\sc i}] emission, as little or no neutral oxygen exists in the ionized flow. \citet{alexander08b} subsequently made similar calculations for the [Ne\,{\sc ii}] 12.81$\mu$m line. The high ionization potential of Ne (21.6eV) means that [Ne\,{\sc ii}] emission can only arise in photoionized gas \citep{glassgold07}, and the high critical density (5\,$\times$\,$10^5$cm$^{-3}$) means that most of the emission comes from close to the base of the photoevaporative flow. The predicted line luminosity is a few $\times$\,$10^{-6}$L$_{\odot}$, and scales linearly with the ionizing luminosity $\Phi$. Moreover, the predicted [Ne\,{\sc ii}] line profile varies strongly with disk inclination: for edge-on disks the (Keplerian) rotation dominates, leading to broad, double-peaked lines, while for face-on disks the line is narrower ($\simeq$\,10km\,s$^{-1}$) and blue-shifted by 5--7km\,s$^{-1}$ (see also Fig.\ref{fig:neii}). \citet{alexander08b} noted that detection of this blue-shift is possible with current mid-IR spectrographs (i.e., at $\lambda/\Delta\lambda$\,$\simeq$\,30,000), and would represent a clear signature of a low-velocity, ionized wind.

\citet{glassgold07} showed that disk irradiation by stellar X-rays can result in strong [Ne\,{\sc ii}] (and [Ne\,{\sc iii}]) emission from the disk surface layers. \citet{hg09} subsequently presented analytic calculations of a number of fine-structure and hydrogen recombination lines from the EUV and X-ray heated layers. They found that the IR fine-structure lines scale linearly with the EUV and X-ray luminosities, and suggested that the ratios of the [Ne\,{\sc ii}] 12.81$\mu$m, [Ne\,{\sc iii}] 15.55$\mu$m and [Ar\,{\sc ii}] 6.99$\mu$m lines can be used as diagnostics of the incident spectrum, potentially distinguishing between EUV and X-ray irradiation. By comparing with the available data (from {\it Spitzer}) they concluded that internal shocks (in jets) or X-ray excitation dominates the production of the [Ne\,{\sc ii}] line, unless the incident EUV spectrum is very soft. \citet{eo10} then combined Monte Carlo radiative transfer calculations with the hydrodynamic wind solution of \citet{owen10} to construct a detailed atlas of emission lines from X-ray-heated winds. They computed fluxes and profiles for almost 100 lines, looking at both collisionally-excited lines from metals and recombination lines of H/He, and investigated how the line emission varied with inclination and $L_{\mathrm X}$. \citet{eo10} again found blue-shifted [Ne\,{\sc ii}] emission (detected by \citealt{ps09}) to be the ``smoking gun'' of disk photoevaporation (see Fig.\,\ref{fig:neii}), but the relatively low ionization fraction ($\chi_{\mathrm e}$\,$\simeq$\,0.01) of the X-ray-heated wind means that, despite the much larger wind rate, the predicted [Ne\,{\sc ii}] luminosity and line profile are both very similar to those predicted for an EUV-driven wind (which has $\chi_{\mathrm e}$\,$\simeq$\,1). \citet{eo10} were also able to reproduce the observed low-velocity component of the [O\,{\sc i}] 6300\AA\ line, which they found to be excited by collisions with neutral hydrogen in the X-ray-heated wind. \citet{gorti11} also find that the [Ne\,{\sc ii}] emission is likely to originate in a predominantly neutral, X-ray-heated region but, for the specific case of TW Hya (where the [O\,{\sc i}] 6300\AA\ line is not blue-shifted; see Section \ref{sec:obs_twhya}), \citet{gorti11} instead argue that the observed [O\,{\sc i}] emission is primarily non-thermal, and find that OH photodissociation (by stellar FUV photons) in a bound disk layer at $\sim$\,10AU naturally reproduces the luminosity and profile of the [O\,{\sc i}] line \citep[see also][]{rigliaco13}. Taken together, these results suggest that observations of line emission should allow us to distinguish between different photoevaporation models.

The simplest statistical predictions of disk dispersal theory are the evolutionary time-scales. The predictions of photoevaporative clearing models are generic in this respect: all predict rapid inside-out disk dispersal after $\sim$Myr disk lifetimes, satisfying the two-time-scale condition discussed in Section \ref{sec:intro_obs}. The absolute time-scales primarily reflect the choice of disk model (particularly the viscosity and initial mass distribution), but the relative time-scales (e.g., the ratio of the clearing time to the lifetime) are only weakly dependent on these parameters \citep{clarke01,alexander06b,alexander08a}. The main difference between the models is that the clearing time is longer in the X-ray and FUV cases than the EUV \citep[e.g.,][see also Fig.\ref{fig:mdot_time}]{gorti09,owen10}. The larger critical radius in these models results in a longer viscous time-scale at the gap-opening radius, and the higher wind rates mean that disk clearing begins at a higher disk accretion rate. It therefore takes somewhat longer for the inner disk to drain on to the star (this phase was termed ``photoevaporation-starved accretion'' by \citealt{drake09} and \citealt{owen11}) and, because the disk mass is higher when the gap opens, photoevaporation also takes longer to remove the outer disk.

The most robust statistical prediction these models make is the distribution of disk accretion rates. In viscous disk models the accretion rate $\dot{M}(t)$ declines as a power-law \citep[e.g.,][]{lbp74,hartmann98}, and the photoevaporative wind essentially sets a lower cut-off to this power-law (as seen in Fig.\ref{fig:mdot_time}). We therefore expect a power-law distribution of accretion rates, truncated below the wind rate $\dot{M}_{\mathrm w}$. \citet{aa09} showed that, for EUV photoevaporation, a modest spread in disk parameters broadly reproduces both the magnitude and scatter of observed accretion rates. X-ray photoevaporation rates scale with the X-ray luminosity (Equation \ref{eq:mdot_X}), and consequently the statistical models of \citet{owen11} found a negative correlation between accretion rate $\dot{M}$ and $L_{\mathrm X}$ (in a co-eval population). We can make similar predictions for the distribution of disk masses, but the model disk masses are inevitably degenerate with the magnitude of the disk viscosity (as $\dot{M}$\,$=$\,$3 \pi \nu \Sigma$ in a steady-state accretion disk). For $\alpha$\,$\sim$\,0.01 and EUV photoevaporation, the disk mass at the start of the clearing phase is $\sim$\,0.001\Msun\ \citep{clarke01,alexander06b}. The higher wind rates in X-ray and FUV models result in larger disk masses before disk clearing begins \citep{gorti09,owen10}, and these models generally favour lower disk viscosities in order to reproduce observed disk lifetimes and masses.

Moving beyond global disk properties, \citet{alexander06b} used a simple prescription to model the SEDs of their evolving disks. They showed that their models were consistent with the observed spread of CTT fluxes across a wide range of wavelengths (from near-IR to sub-mm) and, crucially, showed that the rapid clearing phase of the evolution corresponds to the poorly-populated region between the observed loci of CTTs and WTTs (see Fig.\ref{fig:IR_mm}). \citet{aa09} found that a modest spread in disk parameters results in a significant scatter in disk lifetimes, and that the disk fraction of the population declines in a manner consistent with observations \citep[e.g.,][]{mamajek09}. The X-ray models of \citet{owen11} also successfully reproduce the observed decline in disk fraction with age. However, in this case the clearing time-scale is significantly longer than in the EUV models, as discussed above, and consequently \citet{owen11} also predicted a significant population of non-accreting disks with large inner holes, which should only be detected at $\gtrsim$\,50$\mu$m (though thermal sweeping may disperse these disks rapidly).

Finally, we consider the predicted properties of so-called transitional disks. Broadly speaking, these are objects which are observed to have properties between those typical of CTTs and WTTs (see Section \ref{sec:obs_tds} and the chapter by {\em Espaillat et al.}~for more details), and most observational definitions of ``transitional'' require some degree of inner dust disk depletion \citep[e.g.,][]{strom89}. \citet{alexander06b} noted that the inside-out clearing characteristic of UV-switch models invariably gives rise to a short ``inner hole'' phase, and suggested that some subset of the known transitional disks may be undergoing photoevaporative clearing. The predicted properties of such objects are again fairly generic: an inner disk cavity with size $\gtrsim R_{\mathrm c}$; little or no on-going accretion ($\dot{M}$\,$\lesssim$\,$\dot{M}_{\mathrm w}$); and a small outer disk mass. However, a number of other mechanisms have been also proposed to explain the observed transitional disks, ranging from the dynamical influence of planets to the evolution and growth of small dust grains \citep[e.g.,][]{rice03,quillen04,dd05,cmc07,krauss07}. \citet{aa07} noted that the properties of planet-cleared inner disks are distinct from those cleared by photoevaporation, and suggested that the distribution of transitional disks in the $\dot{M}$--$M_{\mathrm d}$ plane can be used to distinguish between different mechanisms for producing disk inner holes (see also \citealt{najita07}, and the reviews by \citealt{najita_ppv} and \citealt{alexander08a}). \citet{aa09} showed that both of these mechanisms can operate simultaneously, and noted that the observational definition of ``transitional'' has a strong influence on how samples of such objects are interpreted \cite[see also][]{alexander08a}. \citet{owen11} found that many of the observed transitional disks are consistent with models of X-ray photoevaporation, and suggested that the distribution of transitional disks in the $\dot{M}$--$R_{\mathrm {hole}}$ plane also represents an important diagnostic. \citet{owen12} also find that if the process of ``thermal sweeping'' operates as predicted, then outer disk clearing is extremely rapid and the population of non-accreting transitional disks should be relatively small. Interpreting observations of transitional disks remains challenging, but these results show that the properties of transitional disks can offer important insights into the process(es) of disk dispersal.


\subsection{MHD disk winds}\label{sec:models_mhd}
\noindent \citet{blandford82} showed that an accretion disk threaded by a poloidal magnetic field can drive a magnetohydrodynamic (MHD) outflow. Unlike thermally driven winds, MHD outflows remove mass while also exerting a torque on the disk surface, and hence have a qualitatively different impact on secular disk evolution (see the chapters by {\em Frank et al.} and {\em Turner et al.}). MHD winds can result in rapid disk evolution, and could even preclude disk formation entirely \citep[][see also the chapter by {\em Li et al.}]{li11}. Stars with Myr-old disks evidently avoided that fate, perhaps as a consequence of having formed from gas with a relatively weak field \citep{krumholz13}, but the remnant flux could still be dynamically important over longer time scales. The physics of ``wind-driven" disks has been considered in detail by \citet{salmeron11}, and MHD effects could contribute to disk dispersal if the typical magnetic flux threading protostellar disks is sufficiently large. 

An organized magnetic field that supports a wind is more effective at driving disk evolution [by a factor $\sim$\,$(H/R)^{-1}$] than a turbulent field of the same strength. In general, however, there is no reason why winds and turbulent transport cannot coexist \citep{balbus98,shu07}. \citet{suzuki09} simulated the evolution of the magnetorotational instability (MRI) in local stratified domains, threaded by a vertical field with a mid-plane plasma $\beta$\,$=$\,$8 \pi \rho c_s^2 / B_z^2$ that varied between $10^4$ and $10^7$. In their fiducial case, with $\beta$\,$=$\,$10^6$ (physically, a vertical field of $\simeq$\,$10^{-2}$G at 1AU), a few percent of the disk mass was lost within a hundred orbits, and these mass-loss rates are easily large enough to be important for disk dispersal \citep{suzuki10}.

Outflows have been observed to form robustly in local disk simulations whenever a vertical field is present. \citet{fromang13} greatly extended the work of \citet{suzuki09}, and studied how the derived outflows depended on critical numerical parameters (the vertical domain size, boundary conditions, and resolution). Mass loss was observed, but angular momentum transport (for $\beta$\,$=$\,$10^4$) was dominated by turbulent rather than wind stresses. \citet{bai13b,bai13} studied both the ideal-MHD limit and the specific case of protoplanetary disks, simulating the MRI in a local domain with vertical profiles of Ohmic and ambipolar diffusion appropriate to conditions at 1AU. Mass loss was again observed, but in this dead zone regime \citep{gammie96} the wind also dominated the evolution of angular momentum.

The existing simulations exhibit an unphysical dependence of the outlow properties on the boundary conditions, which is likely to be associated with the inherent approximation of a local geometry \citep{lesur13}. Estimated mass-loss rates are as high as $\sim$\,$10^{-8}$\Msunyr\ \citep{bai13c,simon13}, but global simulations are required to quantify the true mass loss rate and to make testable observational predictions. Nonetheless, it seems clear that there is a continuum between the classical limits of viscous and wind-driven disks, and that as mass is accreted the dynamical importance of any non-zero vertical flux must rise, potentially becoming important during the dispersal phase \citep{armitage13}.

\bigskip

\section{OBSERVATIONS OF DISK DISPERSAL}\label{sec:obs}
\noindent Having outlined the theory of disk dispersal, we now consider the observational evidence, focusing in particular on observations which inform and constrain our theoretical models. We consider the high-energy radiation fields of TTs, direct diagnostics of disk photoevaporation, and indirect, statistical studies of disk evolution and dispersal. We also review what we can learn from observations of transitional disks, before presenting a detailed case study of TW Hya (our nearest and best-studied protoplanetary disk).


\subsection{High-energy emission from T Tauri stars}\label{sec:rad_fields}
\noindent As discussed in Section \ref{sec:models_evap}, the input radiation fields remain a major uncertainty in models of disk photoevaporation. This problem is particularly acute in the EUV, where interstellar absorption prohibits direct observation of ionizing photons. \citet{ks04} used a scaling argument (based on solar observations) to estimate the high-energy spectrum of a young, active G-type star, and their spectrum has an ionizing luminosity $\Phi$\,$\simeq$\,$2.5\times10^{41}$s$^{-1}$. \citet{acp05} then used previously derived emission measures to estimate the ionizing emission from several massive, luminous CTTs, and derived values $\Phi$\,$\gtrsim$\,$10^{42}$s$^{-1}$. \citet{acp05} also suggested that the UV He\,{\sc ii}/C\,{\sc iv} line ratio may be used as a diagnostic of the chromospheric emission from TTs; however, recent high-resolution spectra show that for CTTs these lines in fact originate primarily in the accretion flow, and do not trace the chromospheric emission well \citep{ardila13}. \citet{herczeg07b} estimated $\Phi$\,$\simeq$\,$5$\,$\times$\,$10^{41}$s$^{-1}$ for TW Hya, and \citet{espaillat13} estimated $\Phi$\,$\simeq$\,$10^{42}$--$10^{43}$s$^{-1}$ for SZ Cha, but accurate measurements of the EUV luminosity remain scarce (though free-free emission offers a promising alternative diagnostic, as discussed below).

By contrast, X-rays from TTs have been observed for more than 30 years \citep[e.g.,][]{fd81}, and are now well-characterised \citep{feigelson07,gn09}. X-ray luminosities range from $L_{\mathrm X} $\,$\lesssim$\,$10^{28}$erg\,s$^{-1}$ to $L_{\mathrm X} $\,$\gtrsim$\,$10^{32}$erg\,s$^{-1}$, with a spectrum that peaks around 1keV and is broadly consistent with emission from a $\sim$\,$10^7$K plasma. Some fraction of the observed X-ray emission (particularly at low energies) may originate in the accretion flow \citep{kastner02,dupree12}, and this poorly characterised ``soft excess'' (at 0.3--0.4keV) dominates the X-ray luminosity of a small number of CTTs \citep{gn09}. However, magnetic reconnection events in the stellar chromosphere and/or corona are thought to dominate the X-ray emission from the majority of TTs \citep[e.g.,][]{fm99}. X-ray emission from TTs is highly variable, and shows a weak anti-correlation with measured accretion rates \citep{feigelson07}, but young stars show only a modest decline in their X-ray emission on time-scales $\sim$100Myr \citep{ingleby11,stelzer13}.

FUV observations of TTs are more difficult to interpret, as the bulk of the FUV emission from CTTs originates in the accretion shock and to first order $L_{\mathrm {FUV}}$\,$\propto$\,$\dot{M}$ \citep[e.g.,][]{calvet98,gullbring98,yang12}. However, for high $\dot{M}$ the accretion columns and any magnetically-driven jet or outflow strongly shield the disk from UV photons produced in the accretion shock, and in the models of \citet{gorti09} the photoevaporative mass-loss rate is in fact only weakly dependent on the stellar accretion rate.  As discussed in Section \ref{sec:models_evo}, however, one cannot self-consistently use the accretion luminosity to shut off disk accretion, so the chromospheric FUV emission is most critical for disk dispersal. Recent observations by \citet{ingleby12} find that the chromospheric FUV in the range 1230--1800\AA\ ($h\nu$\,$\simeq$\,7--10eV) saturates at $\simeq$\,$10^{-4}$L$_*$.  However, \citet{france12} and \citet[][see also \citealt{herczeg02,herczeg04}]{schindhelm12} were recently able to reconstruct the FUV radiation fields of several CTTs from spectra of fluorescent H$_2$ emission, and found that the FUV luminosity is dominated by line emission, with $\gtrsim$\,90\% of the total FUV flux being emitted in Ly$\alpha$.  The integrated FUV luminosity is therefore $L_{\mathrm {FUV}}$\,$\sim$\,$10^{-3}$L$_*$. \citet{gorti09} adopt a constant stellar/chromospheric luminosity of $L_{\mathrm {FUV}}$\,$\simeq$\,$5\times10^{-4}$L$_*$, consistent with these observations, but the models do not yet include the large contribution from Ly$\alpha$. This is unlikely to alter the heating rates dramatically, but should be taken into account in future studies.

HJLS94 and \citet{lugo04} computed the free-free emission from photoevaporative winds around massive stars, and \citet{pascucci12} have recently calculated the free-free continuum emission and H radio recombination lines arising from a fully- (EUV) or partially-ionized (X-ray) protoplanetary disk surface. They show that the free-free continuum produces excess emission on top of the dust continuum at cm wavelengths, and is detectable with current radio instruments. Such excess emission at 3.5cm is detected from the photoevaporating disk around TW Hya \citep[][and references therein]{wilner05}. \citet{pascucci12} show that if the stellar X-ray luminosity is known, one can estimate the X-ray contribution to the free-free emission and thus find the EUV contribution; in other words, it is possible to measure the stellar EUV flux that the disk receives. In the case of TW Hya, \citet{pascucci12} find that EUV photons dominate the observed free-free emission and estimate $\Phi$\,$\sim$\,$5$\,$\times$\,$10^{40}$s$^{-1}$ at the disk surface. \citet{owen13} have recently extended this analysis with detailed numerical calculations of free-free emission from EUV- and X-ray-irradiated disks, and find that the free-free emission scales approximately linearly with either $\Phi$ or $L_{\mathrm X}$, in agreement with \citet{pascucci12}. \citet{owen13} also argue that if disks can be observed close to the end of their lifetimes (i.e., where photoevaporation starts to overcome disk accretion), then the free-free flux should scale $\propto$\,$\dot{M}^2$ in the EUV-driven case, but $\propto$\,$\dot{M}$ in the X-ray driven case.


\subsection{Direct observations}\label{sec:obs_direct}
\noindent As discussed in Section \ref{sec:models_pred}, directly probing photoevaporative flows requires us to identify the gas lines which trace the heated disk surface layers. {\it Spitzer} Infrared Spectrograph (IRS) observations were the first to discover such possible tracers, via the \neii{} emission line at 12.81$\mu$m \citep{pascucci07,lahuis07}. Due to the low spatial and spectral resolution of the {\it Spitzer} IRS, these data cannot prove that the \neii\ line is indeed a disk diagnostic for individual sources. However, studies of over 100 TTs show that sources with known jets/outflows have systematically higher \neii\ luminosities (by 1--2 orders of magnitude) than sources with no jets, and also find a weak correlation between $L_{\mathrm {[Ne\,II]}}$ and $L_{\mathrm X}$ \citep{gudel10,baldovin12}. These results point to shock-induced emission in circumstellar gas dominating the {\it Spitzer} \neii{} fluxes of jet sources, but lend support to the disk origin for evolved and transitional disks with no jets. \citet{szulagyi12} subsequently considered the detection statistics of the \neii\ 12.81$\mu$m, \neiii{} 15.55$\mu$m, and \arii{} 6.98$\mu$m lines in a large sample of transitional disks and measured the line flux ratios. Although the number of detections is small, the \neii\ line is typically 10 times brighter than the \neiii\ line, and similar in flux to the \arii{} line. Charge exchange between Ne$^{++}$ and H in partially ionized gas naturally leads to a [Ne\,{\sc ii}]/[Ne\,{\sc iii}] ratio $\sim$10 \citep{glassgold07,hg09}, which suggests that X-ray irradiation dominates the heating and ionization of the disk surface traced by these forbidden lines (though [Ne\,{\sc ii}]/[Ne\,{\sc iii}] ratios less than unity have recently been reported for some Class I \& II sources; \citealt{espaillat13,kruger13}).


\begin{figure}[t]
 \centering
 \includegraphics[angle=270,width=\hsize]{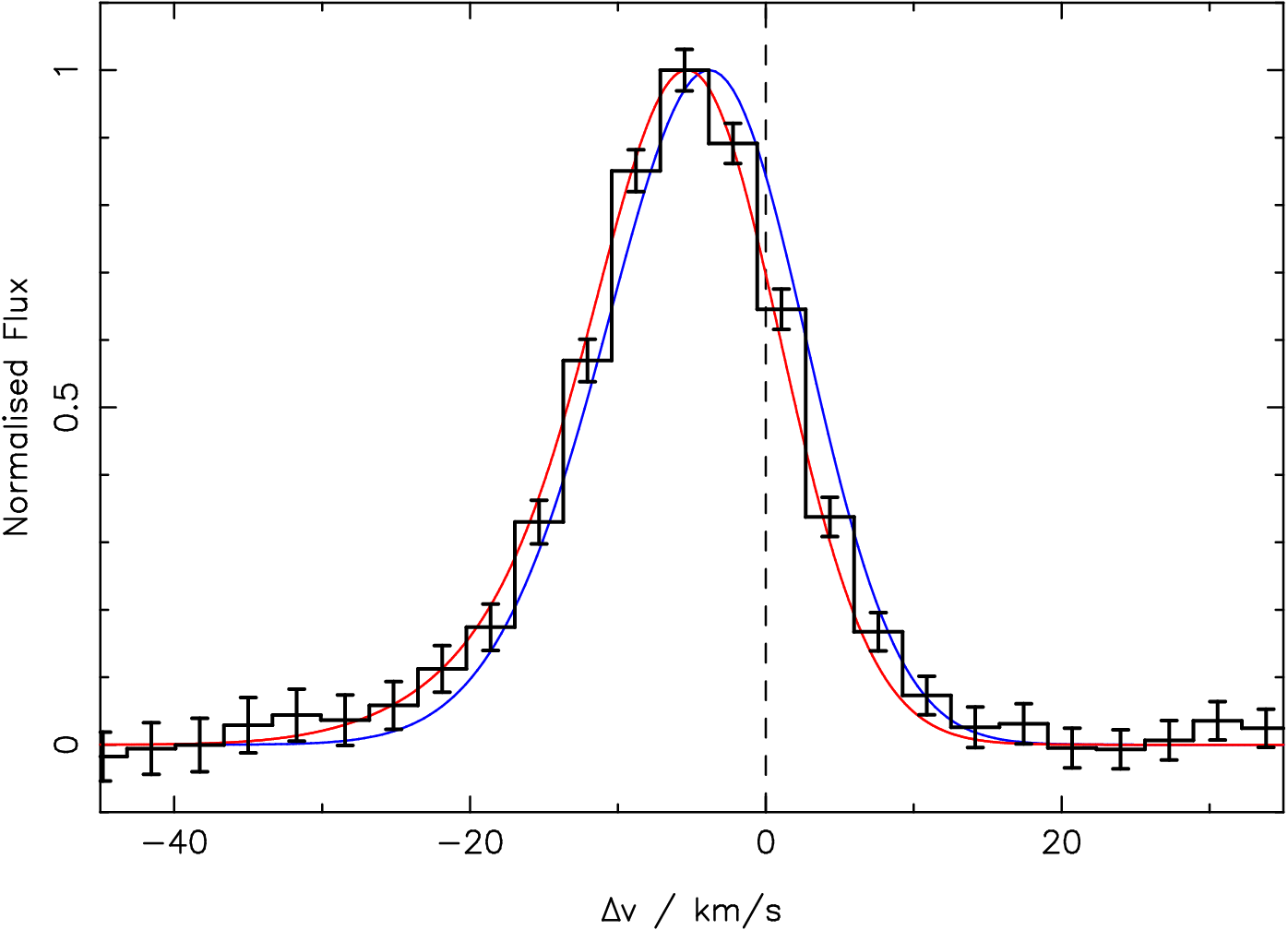}
\caption{\small \neii{} 12.81$\mu$m line profile from the near face-on disk around TW Hya. The black line (and error bars) show the observed profile from \citet{pascucci11}, obtained by co-adding the spectra from the different position angles at which they observed. The red curve shows the theoretical prediction for an EUV-driven wind \citep{alexander08b}, while the blue curve shows the corresponding prediction for X-ray photoevaporation \citep{eo10}; both profiles have been normalised by their peak flux. The observed blue-shift of $5.4\pm0.6$km\,s$^{-1}$ represents an unambiguous detection of a slow, ionized wind, but \neii{} observations alone do not distinguish between the different models.}
 \label{fig:neii}
\end{figure}


As the line falls in an atmospheric window, bright \neii{} emission can be followed up with high-resolution ($\sim$10\kms) ground-based spectrographs, allowing us to disentangle the disk from the jet/outflow contribution. The first such observations hinted at a flux enhancement on the blue side of the \neii{} line from the transitional disk TW Hya \citep{herczeg07}. Higher signal-to-noise spectra then found unequivocal evidence of central star-driven photoevaporation in three transitional disks (TW Hya, T Cha and CS Cha): modest line-widths (FWHM\,$\simeq$\,15--40km\,s$^{-1}$) accompanied by small ($\simeq$\,3--6\kms) blue-shifts in the peak centroid \citep[see also Fig.\ref{fig:neii} and Section \ref{sec:obs_twhya}]{ps09}. At the time of writing 55 {\it Spitzer} \neii{} detections have been followed up with ground-based high-resolution spectrographs \citep{vanboekel09,najita09,ps09,sacco12,baldovin12}. 24 of these resulted in detections, with all the detected lines also being spectrally resolved. Eight of these detections are Class~I/II sources where most of the unresolved {\it Spitzer} \neii{} emission clearly arises in jets/outflows: the \neii{} emission is broad ($\ge$40\kms) and blueshifted by $\ge$\,50\kms\ \citep[and in the case of T Tau is also spatially resolved;][]{vanboekel09}. The \neii{} lines from three sources (AA Tau, CoKuTau/1 and GM Aur) were interpreted as tracing bound gas in a disk, heated by stellar X-rays \citep{najita09}. However, AA Tau and CoKuTau/1 are seen close to edge-on ($i $\,$>$\,70$^\circ$) and are known to power jets \citep[e.g.,][]{heg95,baldovin12}, which may contaminate the observed \neii{} lines. Finally, 13 sources have narrow \neii{} lines ($\sim$\,15--50\kms) and small blue-shifts ($\sim$\,2--18\kms), consistent with photoevaporative winds. Among these objects the transitional disks typically show smaller line-widths and blue-shifts than the Class~I/II sources, but the interpretation of these wind sources is not straightforward. \neii{} line-widths of 15--25\kms\ can be produced in photoevaporative winds \citep{alexander08b,eo10}, but broader lines are difficult to reconcile with photoevaporation models unless the disks are viewed close to edge-on. In addition, the measured blue-shifts cluster around $\simeq$\,10\kms\ \citep{sacco12}, somewhat larger than predicted for X-ray winds and more in line with EUV-driven wind models \citep{alexander08b}. Thus, while small blue-shifts in the \neii{} emission unambiguously point to on-going photoevaporation, \neii{} lines alone cannot be used to measure photoevaporation rates \citep{pascucci11}.  This is due to the degeneracy discussed in Section \ref{sec:models_pred}: a low-density wind with a high ionization fraction (as predicted for EUV photoevaporation) and a higher-density wind with a lower ionization fraction (as predicted for X-ray photoevaporation) both result in very similar \neii{} emission. Further diagnostics are therefore necessary to determine the primary heating/ionization mechanism, and to measure photoevaporative wind rates.

These results, and the predictions of photoevaporative wind models, have recently motivated a re-analysis of the optical forbidden lines detected toward TTs. In particular, oxygen forbidden lines have long been known to display two components: a high-velocity component (HVC), blue-shifted by hundreds of \kms\ with respect to the stellar velocity; and a low velocity component (LVC), blue-shifted by a few to several \kms\ \citep[e.g.,][]{heg95}. While the HVC unambiguously traces accretion-driven jets, as with the \neii{} HVC, the origin of the LVC has remained a mystery. As discussed in Section \ref{sec:models_pred}, reproducing the large \oi\ line luminosities via thermal excitation in a wind requires a mostly neutral layer at high temperatures ($\sim$\,8,000K), as predicted for soft X-ray heating \citep{font04,hg09,eo10}. Alternatively, the \oi{} LVC could trace a cooler ($<$1,000K) disk layer where neutral oxygen is produced by OH photodissociation, as proposed for TW Hya by \citet{gorti11}. In this case the observed \oi{} emission is not blue-shifted (somewhat unusually), hinting at a disk rather than wind origin \citep{pascucci11}. Comparison of \neii{} and \oi{} line profiles in this manner is still limited to a small sample. However, in the five sources where contamination from jet emission can be excluded the \neii{} line shows a larger peak blue-shift than the \oi{} line, and the line profiles are sufficiently different to suggest that the two lines originate in physically distinct regions \citep{pascucci11,rigliaco13}. \citet{rigliaco13} also re-analyzed the Taurus TT sample of \citet{heg95} and found: i) a tight correlation between the luminosity of the \oi\ LVC and the stellar accretion rate (and therefore the stellar FUV flux); ii) a relatively small range of \oi\,6300/5577\AA\ line ratios over a very large range in luminosity, which they argue is difficult to reproduce in thermally-heated gas \citep[see also][]{gorti11}. These results suggest that the \oi\ LVC traces the region where stellar FUV photons dissociate OH molecules, and the typical blue-shifts ($\sim$\,5\kms) point to the emission arising in unbound gas. Whether this wind is FUV-, X-ray- or magnetically-driven remains unclear. However, if the \oi\ LVCs do trace photoevaporative winds then photoevaporation must be ubiquitous in Class II disks, with mass-loss rates $\dot{M}_{\mathrm w}$\,$\gtrsim$\,$10^{-9}$\Msunyr.

Finally, large disk surveys of ro-vibrational CO line emission at 4.7$\mu$m have recently identified an interesting sub-class of single-peaked CO line sources \citep[e.g.,][see also the chapter by {\em Pontoppidan et al.}]{brown13}. The spectro-astrometric signal of the highest S/N examples is consistent with a combination of gas in Keplerian rotation plus a slow (few \kms) disk wind, at $\sim$\,AU radii \citep{pontoppidan11}. \citet{brown13} also find that the majority of the CO profiles have excess emission on the blue side of the line, which further supports the wind hypothesis and suggests that molecular winds are common in Class I/II sources. Whether these winds are thermally- or magnetically-driven is not clear, but several CO wind sources are also known to have strong \oi\ HVC emission \citep[e.g.,][]{heg95,rigliaco13}. This hints at a magnetic origin, and cold molecular outflows at $\sim$\,AU radii are difficult to reconcile with a purely thermal wind scenario, but detailed comparisons with models have not yet been possible. Future studies of spectrally resolved line profiles should allow us to understand the relationship between these different wind diagnostics, and to measure mass-loss rates empirically.


\subsection{Indirect observations}\label{sec:obs_indirect}
\noindent Much of our knowledge about the evolution and eventual dispersal of circumstellar disk material comes from indirect, demographic studies of the fundamental observational tracers of disk gas and dust. In Section 1 we discussed the basic constraints imposed by these studies. Here we review the observations behind these constraints in more detail, and discuss the extent to which demographic surveys can be used to test theoretical models. The most common approach is to characterize a sample of tracer measurements in a young star cluster of a given age, and then compare that ensemble to similar results obtained for star clusters with different ages: in essence, a relatively straightforward statistical comparison of how the disk tracer varies with time. Alternatively, specific disk dispersal mechanisms can be constrained by studying how these disk tracers vary with respect to environment, stellar host properties, or some other evolutionary proxy. In principle, the relationship between these tracers and age (or its proxy) can then be directly compared with the predictions of disk evolution models.

Arguably the most robust and testable of those predictions is the decay of accretion rates with time demanded by viscous evolution models \citep[e.g.,][see also Section \ref{sec:models_evo}]{hartmann98}. Accretion rates are usually derived from ultraviolet continuum excesses \citep{calvet98,gullbring98} or H recombination lines \citep{muzerolle98,muzerolle01}, benchmarked against magnetospheric flow and accretion shock calculations. Combining these accretion rates with stellar ages, estimated via grids of pre-main-sequence stellar evolution models, there is significant observational evidence to support the standard viscous disk paradigm: both the frequency of accretors and their typical $\dot{M}$ values decrease substantially from $\sim$1 to 10Myr \citep[e.g.,][]{muzerolle00,fang09,sicilia10,fedele10}, though the inferred stellar ages remain subject to significant uncertainties (see the chapter by {\em Soderblom et al.}). However, current surveys of TT accretion rates do not yet allow detailed comparisons with models of disk dispersal; in particular, the lack of useful upper limits for weakly- and non-accreting disks severely limits the statistical power of these data \citep[see, e.g., the discussion in][]{cp06}.

Accretion signatures are definitive evidence for the presence of gas in the inner disk, but the converse is not necessarily true and measured accretion rates do not quantify the available gas mass. However, when grounded on a set of detailed physico-chemical models \citep{gh04,woitke09,woitke10,kamp10,kamp11}, observations of mid-IR (e.g., [S\,{\sc i}], H$_2$) and far-IR (e.g., [O\,{\sc i}], [C\,{\sc ii}]) cooling lines can be sensitive and direct tracers of even small amounts of gas at radii $<$\,50AU. Observations of these lines with {\it Spitzer} \citep[e.g.,][]{hollenbach05,pascucci06,chen06} and {\it Herschel} \citep[e.g.,][]{mathews10,woitke11,lebreton12} indicate that these gas reservoirs are depleted within $\lesssim$10Myr. mm-wave spectroscopic searches for rotational transitions of the abundant CO molecule in older (debris) disks suggest that this depletion time-scale applies at much larger radii as well \citep{zuckerman95,dent95,dent05,najita05}. Additional constraints on the gas mass in the inner disk come from far-UV spectra of H$_2$ electronic transitions \citep[e.g.,][]{lecavelier01,roberge05}, which indicate that stars older than $\sim$10Myr or that exhibit no signatures of accretion have virtually no gas at radii $\lesssim$\,1AU \citep{ingleby09,ingleby12,france12}.

These statistical signatures of gas dispersal in the inner disk are also seen in analogous trends for disk solids. Even a small amount of dust emits a substantial IR continuum luminosity, so determining the presence or absence of warm dust grains in the inner disk (at $\lesssim$\,1AU) is relatively straightforward, even for large samples. IR photometric surveys have been conducted for $\sim$20 nearby young stellar associations, providing estimates of the fraction of young stars with excess emission from warm dust at ages ranging from $<$1 to $\sim$30Myr \citep[e.g.,][]{haisch01,luhman04,lada06,hernandez07}. Summarizing those results, \citet{mamajek09} showed that the inner disk fraction decays exponentially with a characteristic time-scale of $\simeq$\,2.5Myr. As with the gas tracers, there is a substantial population of young stars (up to 40\%\ at $\sim$2Myr) that exhibit no near-IR excess from warm dust in an inner disk, and a small (but not negligible, $\lesssim$10\%) sample of older ($\sim$10--20Myr) pre-main sequence stars that retain their dust (and gas) signatures (e.g., TW Hya; see Section \ref{sec:obs_twhya}). However, recent high-resolution imaging surveys revealed that significant fraction of the youngest ``disk-less'' stars and WTTs are in fact close ($<$\,40AU) binaries \citep{ik08,kraus08,kraus11}. In this case dynamical clearing is expected to suppress most inner disk tracers, and the ``corrected'' disk fraction for single stars in young clusters (with ages $\lesssim$\,1--2Myr) is close to 100\% \citep{kraus12}. Finally, although in most cases there is a correspondence between accretion and dust signatures \citep[e.g.,][]{hartigan90,fedele10}, recent studies have identified a small but significant population of young stars that show weak (or very red) dust emission but no hints of accretion \citep[e.g.,][]{lada06,cieza07,cieza13}: this may be evidence for substantial radial evolution in the disk at late times (see Section \ref{sec:obs_tds}, and the chapter by {\it Espaillat et al}.). 

The luminosity of the optically thin mm-wave emission from a disk is the best available quantitative diagnostic of its dust mass \citep{beckwith90}. However, until recently such observations were difficult to obtain for large samples, so demographic studies are not yet mature. In nearby low-mass clusters with ages of $\sim$1--3Myr, mm-wave emission consistent with a dust mass of 1--1000$M_{\oplus}$ is found for essentially all stars with accretion signatures and near-IR excess emission \citep{aw05,aw07b}. Comparisons with samples in older associations suggest that the typical dust mass has declined substantially by $\sim$5Myr \citep[e.g.,][]{carpenter05,mathews12}. 

These global properties of protoplanetary disks suggest that the accreting stage is mostly driven by viscous evolution, while the non-accreting phase is dominated by a different dispersal process such as photoevaporation \citep[e.g.,][]{wc11}. However, demographic surveys do not currently provide a strong means of discriminating between the theoretical models of disk dispersal discussed in Section \ref{sec:models}. The broad conclusions of early demographic work provided much of the original motivation for these models, but the statistical power of these studies has not advanced significantly in the intervening period. There is a large dispersion in the properties of individual systems at any given stellar age, and most demographic tracers also show a systematic dependence on stellar mass \citep[e.g.,][]{muzerolle05,andrews13,mohanty13}. Moreover, most surveys are biased against disk-less stars and WTTs, and uniform samples of non-detections or upper limits are rare. Environmental factors such as external photoevaporation \citep[e.g.,][see Section \ref{sec:models_ext}]{johnstone98,mann10} or tidal stripping by binary companions \citep[e.g.,][]{al94,harris12} also ``contaminate'' demographic data, further complicating comparisons with models. Thus, while demographic surveys provide important clues to our understanding of disk evolution and dispersal, their ability to discriminate between models is currently limited by a combination of selection biases and poor number statistics. This approach is potentially very powerful, however, and we urge that future demographic surveys strive toward uniform sensitivity in un-biased samples.


\subsection{Transitional disks}\label{sec:obs_tds}
\noindent Broadly speaking, transitional disks are protoplanetary disks with a significant deficit of near-IR and/or mid-IR flux with respect to the median SED of CTTs \citep[e.g.,][]{strom89}. This  definition includes both objects with IR SEDs that are smoothly falling with wavelength, and systems with clear ``dips" in their SEDs. While the former group may contain objects with continuous disks extending inward to the dust destruction radius, disks in the latter group show clear evidence for inner holes and gaps and are the focus of our discussion. (The chapter by {\em Espaillat et al.}~discusses observations of transitional disks in much greater detail.) As discussed in Section \ref{sec:models_pred}, a variety of different mechanisms have been proposed to explain the holes and gaps in transitional disks, including photoevaporation \citep{alexander06b}, grain growth \citep{dd05}, and the dynamical clearing by giant planets or (sub)stellar companions \citep{al94,ld06}. These processes are not mutually exclusive, and are likely to operate simultaneously \citep[e.g.,][]{wc11}. For our purposes, the key question regarding transitional disks is whether or not their gaps and cavities are mainly due to photoevaporation, or to the other processes listed above; essentially, we would like to know what fraction (if any) of the observed transitional disks are undergoing disk dispersal. 

Near-IR interferometry \citep{pott10}, adaptive optics imaging \citep{cieza12}, and aperture masking observations \citep{kraus11} have shown that the observed inner holes and gaps in transitional disks are rarely due to close stellar companions or brown dwarfs. Similarly, grain growth models have difficulties explaining the large (sub)-mm cavities observed in resolved images of transitional disks \citep{birnstiel12}. Moreover, \citet{najita07} and \citet{espaillat12} find that transitional disks tend to have lower accretion rates than CTTs with similar disk masses, which suggests that the radial distribution of gas in transitional disks is different from that in CTTs. Photoevaporaton and dynamical clearing by forming planets thus remain the leading explanations for most inner holes and gaps. In particular, photoevaporative clearing nicely explains the incidence and properties of transitional disks around WTTs. Such systems represent $\sim$10$\%$ of the pre-main-sequence population in nearby molecular clouds \citep{cieza07}, and tend to be very faint at mm wavelengths. Moreover, their SEDs appear to trace the inside-out dispersal of protoplanetary disks, and are consistent with objects seen during passage from the primordial to the debris disk stage \citep{wahhaj10,cieza13}. 

The importance of photoevaporation in accreting transitional disks, however, is more controversial. As discussed in Section \ref{sec:models_pred}, whether or not photoevaporation can produce inner holes and gaps in accreting transitional disks depends on the mass-loss rate, and for sufficiently high rates a gap opens while the disk is still relatively massive. Under these circumstances the inner disk accretes on to the star at detectable levels ($\gtrsim$\,$10^{-10}$\Msunyr) for a significant period of time, and during this phase appears as an accreting transitional disk. The low mass-loss rates predicted for EUV photoevaporation cannot therefore explain observed accreting transitional disks, but the much higher wind rates predicted for X-ray and FUV photoevaporation can account for some accreting transitional objects. However, even these models cannot account for the transitional disks with strong on-going accretion which show very large cavities (R $\gtrsim$ 20--80 AU) in resolved (sub-)mm images \citep{owen11,morishima12}. The masses of the known large-cavity disks are higher than the median disk population but, as (sub-)mm imaging surveys have so far focused on the brightest, most massive disks, the significance of this result is not yet clear \citep{andrews11}. Dynamical clearing (by giant planets) seems to be the most likely explanation for these systems, but even this scenario has difficulty accounting for all of the observed properties of these unusual objects \citep[e.g.,][]{zhu12,co13}.

Transitional disks also provide observational tests for X-ray photoevaporation models, as the mass-loss rates scale with $L_{\mathrm X}$ (Equation \ref{eq:mdot_X}). This implies that i) accreting transitional disks should have, on average, higher $L_{\mathrm X}$ than co-eval CTTs with ``full" disks; ii) there should be a weak anti-correlation between $\dot{M}$ and the size of the cavity, and few objects with large ($\gtrsim$\,20AU) cavities and detectable accretion rates \citep{owen11,owen12}. These theoretical predictions have not been verified in the largest samples of transitional disks studied to date \citep{kim13}. Similarly, observations of protoplanetary disks show no difference between the FUV emission levels of systems with transitional disks and those with full disks \citep{ingleby11}. These results suggest that either the observational uncertainties and/or selection biases are large enough to mask the predicted correlations, or that the inner holes and gaps of accreting transitional disks are mostly not the result of X-ray- or FUV-driven photoevaporation. Higher photoevaporation rates throughout disk lifetimes are also difficult to reconcile with systems that have stringent upper limits for their accretion rates ($<$\,$10^{-10}$\Msunyr), yet show no evidence for holes or gaps in their SEDs \citep{ingleby11a}, and the increasing evidence for non-axisymmetric structures and dynamical clearing in a number accreting transitional disks \citep[e.g.,][]{ki12,cassasus13,vandermarel13} also argues against a photoevaporative origin. Moreover, the (sub-)mm fluxes and accretion rates of transitional disks suggests that there may in fact be two distinct populations of transitional disks \citep{oc12}: those with low disk masses and modest to non-detectable accretion (whose inner holes are primarily due to photoevaporation); and those with large disks masses and high accretion rates (whose inner holes could be caused by giant planet formation). Overall, the properties of transitional disks favour models where photoevaporative mass-loss rates are typically low ($\lesssim$\,$10^{-10}$--$10^{-9}$\Msunyr), and only overcome accretion when disks masses and accretion rates are also low. 


\subsection{TW Hya: a case study of late-stage disk evolution}\label{sec:obs_twhya}
\noindent The transitional disk TW Hya is a unique benchmark for protoplanetary disk physics. In addition to being our nearest \citep[54$\pm$6pc,][]{vanleeuwen07} and best-studied protoplanetary disk, it is one of the oldest known gas-rich systems \citep[$\sim$10Myr,][]{torres08}. The original identification of TW Hya as a transitional object was made by \citet{calvet02}, who found that the observed flux deficit at $\lambda$\,$\le$\,10$\mu$m can be modelled with a disk that is depleted of small ($\lesssim$\,$\mu$m) dust grains within $\lesssim$\,4AU, leaving an optically thin inner cavity. The cavity is not completely empty, however: near- and mid-IR interferometric observations have spatially resolved the warm dust emission, and suggest that an optically thick dust component is present within 4AU \citep{eisner06,ratzka07,akeson11}. The exact location and nature of of this component remain unclear, and both a dust ring \citep{akeson11} and a self-luminous companion \citep{arnold12} have been suggested. Interferometric observations at 7mm find a lack of emission at small radii, which is consistent with the presence of a 4AU dust cavity \citep{hughes07}. The mm-sized dust disk is found to extend out to 60AU, where it appears to be sharply truncated, while the gas component extends to $>$\,200AU \citep{andrews12}.

TW Hya's average accretion rate, obtained from eight different optical diagnostics, is $\simeq$\,$7$\,$\times$\,$10^{-10}$\Msunyr\ \citep{curran11}.  This is an order of magnitude below the median accretion rate of CTTs in Taurus \citep[e.g.,][]{gullbring98}, but still clearly places TW Hya in the group of accreting transitional disks (see Section \ref{sec:obs_tds}). Variability in some of these optical diagnostics points to variable mass accretion \citep[hinting at rates as high as $\sim$\,$10^{-8}$\Msunyr\ at times;][]{ab02} but contamination by a stellar wind may contribute to the observed variations \citep{dupree12}. Reproducing all the observed gas emission lines also requires significant depletion of gas within the 4AU dust cavity, by 1--2 orders of magnitude compared to the gas surface density in the outer disk \citep{gorti11}. Estimates for the outer disk mass range between $5$\,$\times$\,$10^{-4}$\Msun\ \citep{thi10}, suggesting substantial depletion, to 0.06\Msun\ \citep{gorti11}, suggesting little or no depletion with respect to ``normal'' disks. The recent {\it Herschel} detection of HD in the TW Hya disk also favours a large disk mass \citep{bergin13}.

While the presence of a giant planet is the leading hypothesis to explain the 4AU cavity \citep[e.g.,][]{calvet02}, the moderate depletion of dust and gas and the relatively low stellar accretion rate are also consistent with some of the star-driven photoevaporation models discussed in Section \ref{sec:models}. Moreover, the detection of a small blueshift ($\simeq$\,5\kms) in the \neii\ 12.81$\mu$m line demonstrates that the TW Hya disk is indeed losing mass via photoevaporation \citep[][see also Fig.\ref{fig:neii}]{ps09,pascucci11}. In addition, these data show that more than 80\% of the \neii\ emission comes from beyond the dust cavity and is confined within $\sim$10AU \citep{pascucci11}, in agreement with the predictions of EUV- and X-ray-driven photoevaporation models. However, the \oi{} 6300\AA\ line from TW Hya has only a moderate width ($\simeq$\,10\kms) and is centred on the stellar velocity \citep{ab02,pascucci11}, which suggests that the \oi{} emission originates in a bound disk layer rather than a wind \citep{gorti11}. We also note that if TW Hya's excess flux at 3.5cm is due to free-free emission, this implies a EUV luminosity of $\sim$\,$5$\,$\times$$10^{40}$s$^{-1}$ incident on the disk surface \citep{pascucci12}. This is close to the fiducial luminosity assumed in models (Equation \ref{eq:mdot_EUV}), and is similar to the value derived by assuming that the \neii\ emission is entirely due to EUV photoevaporation \citep[$7.5$\,$\times$\,$10^{40}$s$^{-1}$;][]{alexander08b,pascucci11}.

While it is clear that the disk of TW~Hya is currently undergoing photoevaporation, current data are not sufficient to conclude that photoevaporation is responsible for the 4AU cavity. Empirical measurements of the mass-loss rate are needed, and only rates higher than the current accretion rate would be consistent with a cavity carved by photoevaporation. A further drawback of the photoevaporation hypothesis is that there is a relatively short window ($\sim$\,2\,$\times$\,$10^5$yr) during which we could observe a photoevaporation-induced gap and still detect accretion on to the star \citep[e.g.,][]{owen11}. This, coupled with the apparently large disk mass, favours the giant planet hypothesis for clearing the inner disk \citep[e.g.,][]{gorti11}. However, regardless of the which mechanism is ultimately responsible for the formation of the cavity, the disk of TW Hya offers a unique insight into how multiple disk dispersal mechanisms can operate concurrently, and may even couple to one another in driving final disk clearing. As our observational capabilities improve in the coming years we should be able to study many more objects in this level of detail, and build up a comprehensive picture of how disk dispersal operates in a large number of systems.

\bigskip

\section{IMPLICATIONS AND CONSEQUENCES OF DISK DISPERSAL}\label{sec:conseq}
\noindent Having reviewed both the theory and observations of disk evolution and dispersal, we now move on to consider how these processes affect the formation and evolution of planetary systems. Changes in disk properties have the potential to alter the microphysics of planet formation, and the effects of disk dispersal have important consequences for the dynamics of forming planetary systems. Here we consider each of these in turn, before summarizing our conclusions and discussing prospects for future work.


\subsection{Chemical effects and impact on planet formation}
\noindent We have seen that photoevaporative mass loss, whether driven by the central star or by nearby massive stars, removes material from the disk surface over much of the planet-forming epoch. However, for most of the disk lifetime the column directly affected by photoevaporation is only a small fraction of the disk surface density. For typical disk parameters, an X-ray heated surface layer with column density (along the line-of-sight to the star) $N_H$\,$\sim $\,$10^{22}$cm$^{-2}$ reaches down to only 3--4$H$ above the disk midplane at $\sim$AU radii. In such low-density gas the vertical settling time-scale for dust is very short \citep{dd05}, and turbulence will lift only the smallest (sub-$\mu$m) particles \citep{dubrulle95} into the launching zone of the wind. Photoevaporation therefore preferentially removes dust-poor material from the disk, increasing the dust-to-gas ratio as the disk evolves. This latter quantity is known to be crucial in driving collective mechanisms for planetesimal formation, including the streaming and gravitational instabilities \citep[e.g.,][]{chiang10}, and consequently photoevaporation may play a significant role in the formation of planets.

\citet{throop05} studied the effect of photoevaporation on planetesimal formation in the context of externally irradiated protoplanetary disks. They considered disks around low-mass stars, and modelled the dust distribution under the limiting assumption of a nearly laminar disk, where the only source of turbulence is the Kelvin-Helmoltz instability generated when the dust layer becomes too dense \citep{sekiya98}. The disks were then subjected to ``external'' FUV/EUV photoevaporation, as expected close to massive stars in the centre of massive star-forming regions such as Orion (see Section \ref{sec:models_ext}). \citet{throop05} adopted the mass-loss rates of \citet{johnstone98}, and assumed that small dust grains were entrained in the wind as long as their volume density did not exceed the gas density in the wind. Under these conditons, \citet{throop05} found significant photoevaporative enhancement in the dust-to-gas ratio, reaching values high enough to meet the gravitational instability threshold of \citet{youdin02} between 5 and 50AU. 

It is, however, highly unlikely that photoevaporation is a prerequisite for {\em all} planetesimal formation. As \citet{throop05} observed, if planetesimal formation does not begin until the disk dispersal epoch then there is insufficient time to form giant planet cores before all the gas is gone. The most plausible role for photoevaporation in planetesimal formation is instead as a possible mechanism for forming a second generation of planetesimals, at later times or at larger radii than would otherwise be possible. The predicted evolution of the gas surface density during photoevaporation by the central star favors such a scenario. Using a 1-D model of EUV-driven photoevaporation, \citet{aa07} found that radial pressure gradients can lead to the formation of a ring of enhanced dust-to-gas ratio as the gas disk is dispersed from the inside-out. Again, too little gas remains at this stage for the results to be important for gas-giant formation, but planetesimal formation at a late epoch could still play a role in the formation of terrestrial planets or debris disks \citep{wyatt08}. The key uncertainty is how much solid material remains at relatively
large radii late in the disk evolution. Absent an efficient particle trapping mechanism
\citep[e.g.,][]{pinilla12}, the outer region of an evolving disk will be depleted of solids under the action of radial drift long before photoevaporation becomes dominant \citep{tl05,takeuchi05,hughes12}.

Photoevaporation may also affect the chemical evolution of protoplanetary disks. As disks evolve, cool refractory elements condense on to dust grains while volatiles (H and He) primarily remain in the gas phase. As photoevaporation removes mass from the disk surface, the midplane gradually evolves and becomes enriched in refractory elements. This process was invoked by \citet{guillot06} as part of an explanation for the Ar, Kr, and Xe enrichment with respect to H measured in Jupiter's atmosphere by the {\it Galileo} probe \citep{owen99}. \citet{guillot06} considered a disk around a solar-mass star, undergoing viscous evolution and subject to central-star-EUV and external-FUV photoevaporation. In their model hydrogen is lost in the photovaporative wind, while noble gases condense on to grains in the cold outer disk and have a smaller escape rate. The noble gases are then vaporized again in the warmer disk region where Jupiter forms, and delivered to the envelope in the gas phase. The significance of this enrichment process should be re-evaluated in light of the potentially higher X-ray- or FUV-driven photoevaporation rates (see Section \ref{sec:models_evap}), and current thinking as to the efficiency of vertical mixing processes within the disk [\citep{guillot06} assumed that vertical mixing was dominated by convection].


\subsection{Dynamical effects}
\noindent The manner in which protoplanetary disks are dispersed influences the mass and final orbital properties of planets, through its effects on planetary growth, migration, and orbital stability. The rapid decrease in surface density as the disk is dispersed can starve late-forming cores of gas, preventing them growing into fully-formed gas giants. This mechanism was proposed by \citet{shu93} to explain why Saturn, Uranus \& Neptune are gas-poor, with smaller envelopes than Jupiter. The same effect halts inward migration, stranding planets at radii between their formation radius (which may be beyond the snow-line) and the locations of hot Jupiters (with semi-major axis $a$\,$<$0.1AU). If several planets form in close proximity, the removal of gas will stop disk damping of eccentricity and inclination, with further evolution of the system occurring via purely N-body perturbations. These effects are all generic to any disk dispersal mechanism. However, whether they impart identifiable features on the properties of observed planetary systems depends on how quickly, and from which radii, gas is lost during disk dispersal.

Angular momentum exchange between massive planets ($\gtrsim$\,0.5\Mjup) and the protoplanetary gas disk results in Type II migration, in which the planet's orbital evolution within a gap is coupled to the evolution of the disk \citep[e.g.,][]{kley12}. Migration in this regime is typically inward, though outward migration is possible if planets form in a region where the viscous flow of the gas is away from the star, and mass loss from the outer regions of the disk also promotes outward migration \citep{veras04,martin07}. The rate of migration is generally a non-linear function of the local gas disk conditions, and can be estimated in 1-D viscous disk models given knowledge of how angular momentum is transported in the disk \citep{ivanov99}. Disk dispersal inevitably marks the end-point of Type II migration, and hydrodynamic simulations show that migrating giant planets are indeed stranded by final disk dispersal \citep{rosotti13}.

The influence of photoevaporation on the orbital distribution of extrasolar gas-giants was included in early population synthesis calculations by \citet{armitage02}, using a simple analytic prescription for external FUV photoevaporation \citep[see also][]{matsuyama03b}. These models were extended by \citet{aa09}, who studied giant planet migration and the formation of transition disks using a 1-D disk model that included both viscous transport of angular momentum and EUV photoevaporation. \citet{aa09} assumed that the time at which gas-giants form is uniformly distributed toward the end of the disk lifetime, and found that the observed distribution of exoplanet semi-major axis (within a few AU) is consistent with planet formation further out ($\gtrsim$\,5AU), followed by Type II migration and stranding when the disk is dispersed. Integrated over all (giant) planet masses, the distribution depends primarily upon the nature of angular momentum transport in the disk \citep[and hence is modified in the presence of a dead zone, e.g.,][]{armitage07,matsumura09}, but is also affected by uncertainties in the rate of mass and angular momentum accretion across gaps in the disk \citep{ld06}. The integrated distribution is essentially independent of the details of the photoevaporation model, but sensitivity to the disk dispersal mechanism becomes apparent when the distribution of planet semi-major axes is broken down into different mass bins. \citet{ap12} found that variations in the migration rate near the radius where photoevaporation first opens a gap lead to mass-dependent deserts and pile-ups in the planetary distribution. These effects may be observable in the case of EUV photoevaporation, because the photoevaporative gap in this case falls at small radii ($\simeq$\,1--2AU) where most of the observed planets are likely to have migrated, rather than formed in situ.

State-of-the-art population synthesis models incorporate a much broader range of physical processes than just disk evolution and planet migration, including simplified treatments of core formation, Type I migration, and envelope evolution (see the chapter by {\em Benz et al.}). \citet{mordasini12} include simple prescriptions for both external (FUV) and internal (EUV) photoevaporation, and their models were able to reproduce the observed distribution of planets [$f(a,M_p,R_p)$] reasonably well for planets $\gtrsim$\,2R$_\oplus$. However, in these models uncertainties other than those associated with disk dispersal are dominant. The models of \citet{hp12} similarly invoke internal FUV photoevaporation to drive disk dispersal, but find that the strongest features in the resulting planet population are due to changes in the migration rate at specific locations in the disk \citep[so-called ``planet traps''; e.g.,][]{masset06}. It may therefore be difficult to distinguish the effects of photoevaporation from other physical processes.


\begin{figure*}[t]
 \centering
 \includegraphics[width=165mm]{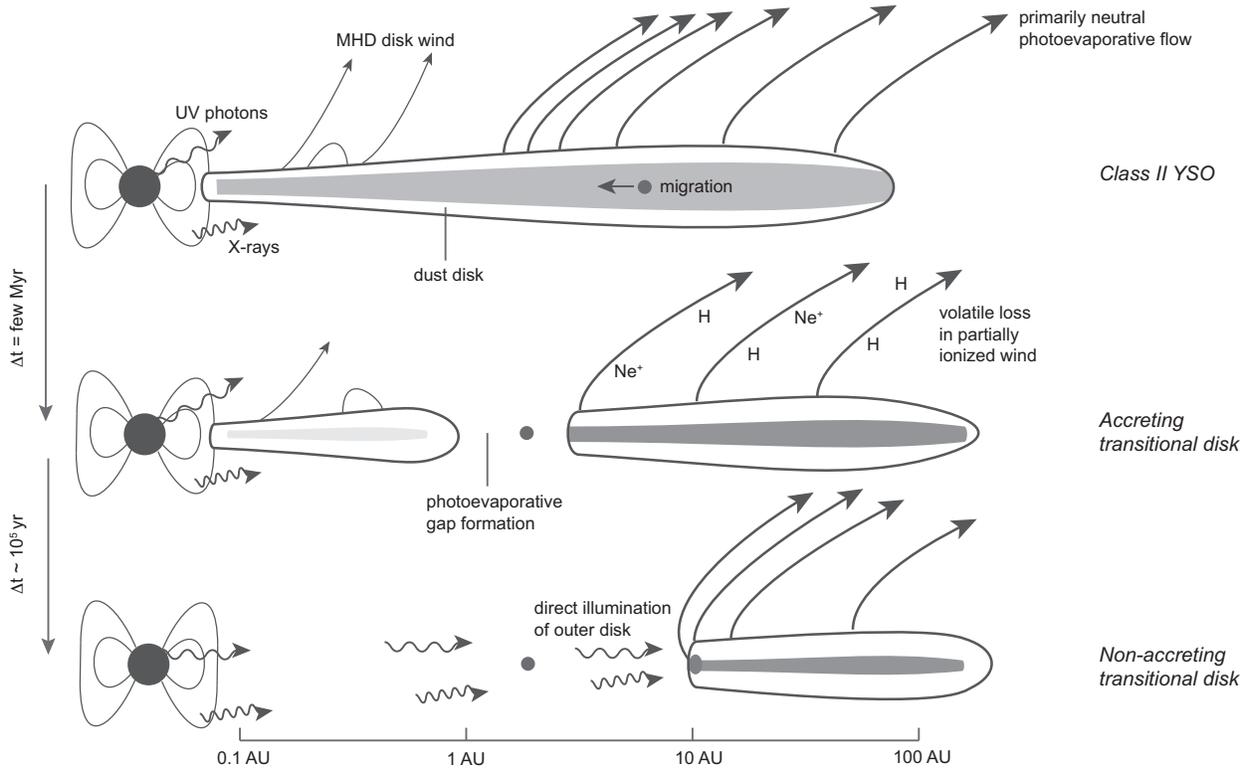}
 \caption{\small Schematic representation of the disk evolution and dispersal sequence outlined in Section \ref{sec:schematic}. At early times accretion dominates the evolution, while mass-loss is dominated by some combination of X-ray and/or FUV photoevaporation and magnetic fields. At later times the photoevaporative flow is at least partially ionized, driven by EUV and/or X-ray irradiation, and once this wind overcomes the accretion flow the disk is rapidly cleared from the inside-out.}
 \label{fig:schematic}
\end{figure*}


Additional dynamical effects arise when multiple planets interact with a dispersing gas disk. In systems with well-separated planets, the changing gravitational potential of the disk during dispersal alters the precession rates of planets and leads to a radial sweeping of secular resonances \citep{nagasaw05}. By contrast, in closely-packed planetary systems gravitational torques between planets and the gas damp eccentricity and inclination \citep{ki02,aw02}, and the presence of a gas disk therefore suppresses planet-planet interactions and scattering. \citet{moeckel12} studied the development of dynamical instabilities during the final phase of X-ray-driven disk clearing, using two-dimensional hydrodynamics to simulate the evolution of a gas disk with three embedded planets. They found that the outcome of dynamical instabilities in the presence of a dispersing disk was similar to gas-free simulations, though a significant number of stable resonant systems formed due to gas-driven orbital migration. However, it remains computationally challenging to model the formation and growth of multiple planets in two-dimensional simulations. As a result it is unclear whether multiple massive planets typically evolve to resonant, packed and rapidly unstable configurations when the gas disk is dispersed, or if stable configurations are preferred \citep{marzari10,lega13}. Moreover, dynamical simulations to date have generally focussed on gas-giant planets, which migrate in the Type II regime. The effects of disk dispersal on the migration and dynamics of lower-mass planets are potentially more significant, but remain largely unexplored by current models.
 
Finally, we note that disks in binary systems may also represent an interesting test of disk dispersal theory. Disk masses are systematically lower in binary systems than for single stars, and disk formation is apparently strongly suppressed around binaries with small ($\lesssim$\,40AU) separations \citep{harris12,kraus12}. However, long-lived circumbinary disks do exist \citep[e.g.,][]{ik08,rosenfeld12}, and the recent discovery that circumbinary planets are relatively common \citep{doyle11,welsh12} has revived interest in the evolution and dispersal of disks around young binary stars. \citet{alexander12} constructed 1-D models of circumbinary disk evolution, and found that the suppression of disk accretion by the tidal torque from the binary greatly enhances the role played by photoevaporation. These results suggests that circumbinary disks may provide a useful laboratory for studying disk dispersal.


\subsection{Schematic picture of disk evolution and dispersal}\label{sec:schematic}
\noindent We have reviewed the theory and observations underpinning our understanding of how protoplanetary disks are dispersed. There have been significant advances in this field since {\em Protostars and Planets V}, and we now have a robust theoretical framework which we are beginning to test directly with observations. Though a number of details remain unresolved, the results discussed above allow us to draw several interesting conclusions:
\renewcommand{\labelenumi}{\roman{enumi}}
\begin{enumerate}
\item Protoplanetary disk evolution on $\sim$\,Myr time-scales is primarily driven by accretion, but  winds (both magnetically-launched and photoevaporative) may drive significant mass-loss throughout the lifetimes of many disks.
\item Disk photoevaporation, driven by high-energy radiation from the central star, is now directly detected in a number of systems, and is the most plausible mechanism for gas disk dispersal.
\item Theoretical models predict that photoevaporative mass-loss rates range from $\sim$\,$10^{-10}$--$10^{-8}$\Msunyr. Current models suggest that X-ray- or FUV-heating dominates in typical systems, yielding mass-loss rates towards the upper end of this range.
\item Inferred mass-loss rates, from both direct and indirect observational tracers, are broadly consistent with these predictions (though the lower end of this range is weakly favoured by current demographic data).
\item High-resolution spectroscopy of emission lines offers a direct test of disk wind models, and may allow mass-loss rates to be measured empirically. 
\item Photoevaporation can explain the properties of some, but not all, transitional disks, and it seems likely that multiple disk clearing mechanisms operate concurrently these systems.
\item Disk dispersal ends the epoch of giant planet formation, and can have a strong influence on the architectures of forming planetary systems. 
\item Mass-loss throughout the disk lifetime may also influence (or even trigger) planet formation, by depleting the disk of gas and enhancing the fractional abundances of both dust and heavy elements.
\end{enumerate}
Based on these conclusions, we are able to construct a tentative schematic picture of protoplanetary disk evolution and dispersal, which is illustrated in Fig.\ref{fig:schematic}. The earliest stages of disk evolution (broadly described as the Class I phase) are dominated by infall on to the disk, and accretion is driven primarily by gravitational instabilities. This phase is also characterised by strong jets and outflows, launched from close to the star by magnetic effects, though their significance in terms of the global evolution of the disk is unclear. The disk then evolves towards a more quiescent evolutionary phase, of which Class II sources and CTTs are typical. Here the evolution is primarily driven by disk accretion, but mass-loss in low-velocity winds is also significant in many, perhaps most, systems. At this stage these winds are primarily neutral, and driven by some combination of X-ray and/or FUV photoevaporation and magnetic fields; the dominant mechanism, and the extent to which such winds deplete or truncate the disk, may well vary from disk to disk. Final disk dispersal begins when mass-loss begins to dominate over accretion. The winds are now at least partially ionized (with ionization fraction $\chi_{\mathrm e}$\,$\gtrsim$\,$10^{-2}$), and significant mass-loss is apparently driven by some combination of X-ray and EUV photoevaporation (which again may vary between disks). This evolutionary phase is broadly associated with the transition from Class II to Class III SEDs, but we stress that a variety of different physical processes contribute to the observed properties of individual ``transitional'' disks. Once photoevaporation becomes the major driver of disk evolution the effect is dramatic. The disk gas is rapidly cleared from the inside-out, stranding any migrating planets at their present locations and profoundly altering the spatial distribution of the remaining disk solids. From this point onwards the nascent planetary system is dominated by gravitational and collisional dynamics, and gradually evolves through the debris disk phase to stability as a mature planetary system.

This picture is obviously somewhat idealised, but is now supported by mature theoretical models and observational evidence. Several important uncertainties remain, however, and we conclude by highlighting the most important areas for future progress. In the short term, emission line studies are perhaps the most promising diagnostic, with the potential to provide empirical measurements of disk mass-loss rates in addition to offering precise tests of theoretical models. The evolution of disk dispersal theory also continues apace, and recent developments in our understanding of magnetically-driven winds have the potential to alter this field significantly in the coming years. On longer time-scales we expect new, high-resolution observational facilities and techniques (especially {\it ALMA}) to revolutionise our understanding of protoplanetary disk physics; recent studies of young binaries and tentative detections of forming planets represent only the tip of this coming iceberg. We also continue to extend our understanding of how disk evolution and dispersal influences planet formation and the architectures of planetary systems, and to build links between protoplanetary disks and our rapidly expanding knowledge of exoplanets. The future of this field is therefore bright, and we look forward to discussing a plethora of exciting new developments at {\em Protostars and Planets VII}.


\bigskip

\noindent \textbf{Acknowledgments} We thank Sylvie Cabrit, Cathie Clarke, Alex Dunhill, Suzan Edwards, Barbara Ercolano, Catherine Espaillat, Uma Gorti, Greg Herczeg, David Hollenbach, James Owen, Klaus Pontoppidan and Elisabetta Rigliaco for a number of insightful discussions. We also thank Uma Gorti \& James Owen for providing some of the data used in Figs.\,\ref{fig:mdot_profiles}, \ref{fig:mdot_time} \& \ref{fig:neii}.  We are grateful to the referee, David Hollenbach, and the editor, Kees Dullemond, for their thoughtful and detailed comments.
RA acknowledges support from STFC through an Advanced Fellowship (ST/G00711X/1) and Consolidated Grant ST/K001000/1. IP acknowledges support from an NSF Astronomy \& Astrophysics research grant (AST0908479). PA acknowledges support from NASA's Origins of Solar Systems Program (NNX13AI58G). LC was supported by NASA through the Sagan Fellowship Program, under an award from Caltech.



\bigskip

\begin{thebibliography}{255}
\parskip=0pt \itemsep=0pt \small \baselineskip=11pt
\providecommand{\natexlab}[1]{#1}

\bibitem[\protect\astroncite{\emph{{Adams} et~al.}}{2004}]{adams04}
{Adams} F.~C. et~al. (2004) \emph{\apj}, \emph{611}, 360.

\bibitem[\protect\astroncite{\emph{{Adams} et~al.}}{2006}]{adams06}
{Adams} F.~C. et~al. (2006) \emph{\apj}, \emph{641}, 504.

\bibitem[\protect\astroncite{\emph{{Agnor} and {Ward}}}{2002}]{aw02}
{Agnor} C.~B. and {Ward} W.~R. (2002) \emph{\apj}, \emph{567}, 579.

\bibitem[\protect\astroncite{\emph{{Akeson} et~al.}}{2011}]{akeson11}
{Akeson} R.~L. et~al. (2011) \emph{\apj}, \emph{728}, 96.

\bibitem[\protect\astroncite{\emph{{Alencar} and {Batalha}}}{2002}]{ab02}
{Alencar} S.~H.~P. and {Batalha} C. (2002) \emph{\apj}, \emph{571}, 378.

\bibitem[\protect\astroncite{\emph{{Alexander}}}{2008{\natexlab{a}}}]{alexander08a}
{Alexander} R. (2008{\natexlab{a}}) \emph{\nar}, \emph{52}, 60.

\bibitem[\protect\astroncite{\emph{{Alexander}}}{2012}]{alexander12}
{Alexander} R. (2012) \emph{\apjl}, \emph{757}, L29.

\bibitem[\protect\astroncite{\emph{{Alexander}}}{2008{\natexlab{b}}}]{alexander08b}
{Alexander} R.~D. (2008{\natexlab{b}}) \emph{\mnras}, \emph{391}, L64.

\bibitem[\protect\astroncite{\emph{{Alexander} and {Armitage}}}{2007}]{aa07}
{Alexander} R.~D. and {Armitage} P.~J. (2007) \emph{\mnras}, \emph{375}, 500.

\bibitem[\protect\astroncite{\emph{{Alexander} and {Armitage}}}{2009}]{aa09}
{Alexander} R.~D. and {Armitage} P.~J. (2009) \emph{\apj}, \emph{704}, 989.

\bibitem[\protect\astroncite{\emph{{Alexander} and {Pascucci}}}{2012}]{ap12}
{Alexander} R.~D. and {Pascucci} I. (2012) \emph{\mnras}, \emph{422}, L82.

\bibitem[\protect\astroncite{\emph{{Alexander}
  et~al.}}{2004{\natexlab{a}}}]{alexander04a}
{Alexander} R.~D. et~al. (2004{\natexlab{a}}) \emph{\mnras}, \emph{348}, 879.

\bibitem[\protect\astroncite{\emph{{Alexander}
  et~al.}}{2004{\natexlab{b}}}]{alexander04b}
{Alexander} R.~D. et~al. (2004{\natexlab{b}}) \emph{\mnras}, \emph{354}, 71.

\bibitem[\protect\astroncite{\emph{{Alexander} et~al.}}{2005}]{acp05}
{Alexander} R.~D. et~al. (2005) \emph{\mnras}, \emph{358}, 283.

\bibitem[\protect\astroncite{\emph{{Alexander}
  et~al.}}{2006{\natexlab{a}}}]{alexander06a}
{Alexander} R.~D. et~al. (2006{\natexlab{a}}) \emph{\mnras}, \emph{369}, 216.

\bibitem[\protect\astroncite{\emph{{Alexander}
  et~al.}}{2006{\natexlab{b}}}]{alexander06b}
{Alexander} R.~D. et~al. (2006{\natexlab{b}}) \emph{\mnras}, \emph{369}, 229.

\bibitem[\protect\astroncite{\emph{{Andre} et~al.}}{1993}]{andre93}
{Andre} P. et~al. (1993) \emph{\apj}, \emph{406}, 122.

\bibitem[\protect\astroncite{\emph{{Andrews} and {Williams}}}{2005}]{aw05}
{Andrews} S.~M. and {Williams} J.~P. (2005) \emph{\apj}, \emph{631}, 1134.

\bibitem[\protect\astroncite{\emph{{Andrews} and {Williams}}}{2007}]{aw07b}
{Andrews} S.~M. and {Williams} J.~P. (2007) \emph{\apj}, \emph{671}, 1800.

\bibitem[\protect\astroncite{\emph{{Andrews} et~al.}}{2009}]{andrews09}
{Andrews} S.~M. et~al. (2009) \emph{\apj}, \emph{700}, 1502.

\bibitem[\protect\astroncite{\emph{{Andrews} et~al.}}{2011}]{andrews11}
{Andrews} S.~M. et~al. (2011) \emph{\apj}, \emph{732}, 42.

\bibitem[\protect\astroncite{\emph{{Andrews} et~al.}}{2012}]{andrews12}
{Andrews} S.~M. et~al. (2012) \emph{\apj}, \emph{744}, 162.

\bibitem[\protect\astroncite{\emph{{Andrews} et~al.}}{2013}]{andrews13}
{Andrews} S.~M. et~al. (2013) \emph{\apj}, \emph{771}, 129.

\bibitem[\protect\astroncite{\emph{{Ardila} et~al.}}{2013}]{ardila13}
{Ardila} D.~R. et~al. (2013) \emph{\apjs}, \emph{207}, 1.

\bibitem[\protect\astroncite{\emph{{Armitage}}}{2007}]{armitage07}
{Armitage} P.~J. (2007) \emph{\apj}, \emph{665}, 1381.

\bibitem[\protect\astroncite{\emph{{Armitage}}}{2011}]{armitage11}
{Armitage} P.~J. (2011) \emph{\araa}, \emph{49}, 195.

\bibitem[\protect\astroncite{\emph{{Armitage} et~al.}}{2001}]{armitage01}
{Armitage} P.~J. et~al. (2001) \emph{\mnras}, \emph{324}, 705.

\bibitem[\protect\astroncite{\emph{{Armitage} et~al.}}{2002}]{armitage02}
{Armitage} P.~J. et~al. (2002) \emph{\mnras}, \emph{334}, 248.

\bibitem[\protect\astroncite{\emph{{Armitage} et~al.}}{2013}]{armitage13}
{Armitage} P.~J. et~al. (2013) \emph{\apjl}, \emph{{in press}} (arXiv:1310:6745).

\bibitem[\protect\astroncite{\emph{{Arnold} et~al.}}{2012}]{arnold12}
{Arnold} T.~J. et~al. (2012) \emph{\apj}, \emph{750}, 119.

\bibitem[\protect\astroncite{\emph{{Artymowicz} and {Lubow}}}{1994}]{al94}
{Artymowicz} P. and {Lubow} S.~H. (1994) \emph{\apj}, \emph{421}, 651.

\bibitem[\protect\astroncite{\emph{{Bae} et~al.}}{2013}]{bae13}
{Bae} J. et~al. (2013) \emph{\apj}, \emph{774}, 57.

\bibitem[\protect\astroncite{\emph{{Bai}}}{2013}]{bai13c}
{Bai} X.-N. (2013) \emph{\apj}, \emph{772}, 96.

\bibitem[\protect\astroncite{\emph{{Bai} and
  {Stone}}}{2013{\natexlab{a}}}]{bai13b}
{Bai} X.-N. and {Stone} J.~M. (2013{\natexlab{a}}) \emph{\apj}, \emph{767}, 30.

\bibitem[\protect\astroncite{\emph{{Bai} and
  {Stone}}}{2013{\natexlab{b}}}]{bai13}
{Bai} X.-N. and {Stone} J.~M. (2013{\natexlab{b}}) \emph{\apj}, \emph{769}, 76.

\bibitem[\protect\astroncite{\emph{{Balbus}}}{2011}]{balbus11}
{Balbus} S.~A. (2011) in: \emph{{Physical Processes in Circumstellar Disks
  around Young Stars}}, (edited by P.~J.~V. {Garcia}), pp. 237--282.

\bibitem[\protect\astroncite{\emph{{Balbus} and {Hawley}}}{1998}]{balbus98}
{Balbus} S.~A. and {Hawley} J.~F. (1998) \emph{Reviews of Modern Physics},
  \emph{70}, 1.

\bibitem[\protect\astroncite{\emph{{Baldovin-Saavedra}
  et~al.}}{2012}]{baldovin12}
{Baldovin-Saavedra} C. et~al. (2012) \emph{\aap}, \emph{543}, A30.

\bibitem[\protect\astroncite{\emph{{Bally} and {Scoville}}}{1982}]{bs82}
{Bally} J. and {Scoville} N.~Z. (1982) \emph{\apj}, \emph{255}, 497.

\bibitem[\protect\astroncite{\emph{{Beckwith} et~al.}}{1990}]{beckwith90}
{Beckwith} S.~V.~W. et~al. (1990) \emph{\aj}, \emph{99}, 924.

\bibitem[\protect\astroncite{\emph{{Begelman} et~al.}}{1983}]{begelman83}
{Begelman} M.~C. et~al. (1983) \emph{\apj}, \emph{271}, 70.

\bibitem[\protect\astroncite{\emph{{Bergin} et~al.}}{2013}]{bergin13}
{Bergin} E.~A. et~al. (2013) \emph{\nat}, \emph{493}, 644.

\bibitem[\protect\astroncite{\emph{{Birnstiel} et~al.}}{2012}]{birnstiel12}
{Birnstiel} T. et~al. (2012) \emph{\aap}, \emph{544}, A79.

\bibitem[\protect\astroncite{\emph{{Blandford} and
  {Payne}}}{1982}]{blandford82}
{Blandford} R.~D. and {Payne} D.~G. (1982) \emph{\mnras}, \emph{199}, 883.

\bibitem[\protect\astroncite{\emph{{Brown} et~al.}}{2013}]{brown13}
{Brown} J.~M. et~al. (2013) \emph{\apj}, \emph{770}, 94.

\bibitem[\protect\astroncite{\emph{{Calvet} and {Gullbring}}}{1998}]{calvet98}
{Calvet} N. and {Gullbring} E. (1998) \emph{\apj}, \emph{509}, 802.

\bibitem[\protect\astroncite{\emph{{Calvet} et~al.}}{2002}]{calvet02}
{Calvet} N. et~al. (2002) \emph{\apj}, \emph{568}, 1008.

\bibitem[\protect\astroncite{\emph{{Carpenter} et~al.}}{2005}]{carpenter05}
{Carpenter} J.~M. et~al. (2005) \emph{\aj}, \emph{129}, 1049.

\bibitem[\protect\astroncite{\emph{{Casassus} et~al.}}{2013}]{cassasus13}
{Casassus} S. et~al. (2013) \emph{\nat}, \emph{493}, 191.

\bibitem[\protect\astroncite{\emph{{Chen} et~al.}}{2006}]{chen06}
{Chen} C.~H. et~al. (2006) \emph{\apjs}, \emph{166}, 351.

\bibitem[\protect\astroncite{\emph{{Chiang} and {Murray-Clay}}}{2007}]{cmc07}
{Chiang} E. and {Murray-Clay} R. (2007) \emph{Nature Physics}, \emph{3}, 604.

\bibitem[\protect\astroncite{\emph{{Chiang} and {Youdin}}}{2010}]{chiang10}
{Chiang} E. and {Youdin} A.~N. (2010) \emph{Annual Review of Earth and
  Planetary Sciences}, \emph{38}, 493.

\bibitem[\protect\astroncite{\emph{{Cieza} et~al.}}{2007}]{cieza07}
{Cieza} L. et~al. (2007) \emph{\apj}, \emph{667}, 308.

\bibitem[\protect\astroncite{\emph{{Cieza} et~al.}}{2008}]{cieza08}
{Cieza} L.~A. et~al. (2008) \emph{\apjl}, \emph{686}, L115.

\bibitem[\protect\astroncite{\emph{{Cieza} et~al.}}{2012}]{cieza12}
{Cieza} L.~A. et~al. (2012) \emph{\apj}, \emph{750}, 157.

\bibitem[\protect\astroncite{\emph{{Cieza} et~al.}}{2013}]{cieza13}
{Cieza} L.~A. et~al. (2013) \emph{\apj}, \emph{762}, 100.

\bibitem[\protect\astroncite{\emph{{Clarke}}}{2011}]{clarke11}
{Clarke} C. (2011) in: \emph{Physical Processes in Circumstellar Disks around
  Young Stars}, (edited by P.~J.~V. {Garcia}), pp. 355--418.

\bibitem[\protect\astroncite{\emph{{Clarke}}}{2007}]{clarke07}
{Clarke} C.~J. (2007) \emph{\mnras}, \emph{376}, 1350.

\bibitem[\protect\astroncite{\emph{{Clarke} and {Owen}}}{2013}]{co13}
{Clarke} C.~J. and {Owen} J.~E. (2013) \emph{\mnras}, \emph{433}, L69.

\bibitem[\protect\astroncite{\emph{{Clarke} and {Pringle}}}{2006}]{cp06}
{Clarke} C.~J. and {Pringle} J.~E. (2006) \emph{\mnras}, \emph{370}, L10.

\bibitem[\protect\astroncite{\emph{{Clarke} et~al.}}{2001}]{clarke01}
{Clarke} C.~J. et~al. (2001) \emph{\mnras}, \emph{328}, 485.

\bibitem[\protect\astroncite{\emph{{Curran} et~al.}}{2011}]{curran11}
{Curran} R.~L. et~al. (2011) \emph{\aap}, \emph{526}, A104.

\bibitem[\protect\astroncite{\emph{{Dent} et~al.}}{1995}]{dent95}
{Dent} W.~R.~F. et~al. (1995) \emph{\mnras}, \emph{277}, L25.

\bibitem[\protect\astroncite{\emph{{Dent} et~al.}}{2005}]{dent05}
{Dent} W.~R.~F. et~al. (2005) \emph{\mnras}, \emph{359}, 663.

\bibitem[\protect\astroncite{\emph{{Doyle} et~al.}}{2011}]{doyle11}
{Doyle} L.~R. et~al. (2011) \emph{Science}, \emph{333}, 1602.

\bibitem[\protect\astroncite{\emph{{Drake} et~al.}}{2009}]{drake09}
{Drake} J.~J. et~al. (2009) \emph{\apjl}, \emph{699}, L35.

\bibitem[\protect\astroncite{\emph{{Dubrulle} et~al.}}{1995}]{dubrulle95}
{Dubrulle} B. et~al. (1995) \emph{\icarus}, \emph{114}, 237.

\bibitem[\protect\astroncite{\emph{{Dullemond} and {Dominik}}}{2005}]{dd05}
{Dullemond} C.~P. and {Dominik} C. (2005) \emph{\aap}, \emph{434}, 971.

\bibitem[\protect\astroncite{\emph{{Dullemond} et~al.}}{2007}]{dullemond_ppv}
{Dullemond} C.~P. et~al. (2007) \emph{Protostars and Planets V}, pp. 555--572.

\bibitem[\protect\astroncite{\emph{{Dupree} et~al.}}{2012}]{dupree12}
{Dupree} A.~K. et~al. (2012) \emph{\apj}, \emph{750}, 73.

\bibitem[\protect\astroncite{\emph{{Duvert} et~al.}}{2000}]{duvert00}
{Duvert} G. et~al. (2000) \emph{\aap}, \emph{355}, 165.

\bibitem[\protect\astroncite{\emph{{Eisner} et~al.}}{2006}]{eisner06}
{Eisner} J.~A. et~al. (2006) \emph{\apjl}, \emph{637}, L133.

\bibitem[\protect\astroncite{\emph{{Ercolano} and {Owen}}}{2010}]{eo10}
{Ercolano} B. and {Owen} J.~E. (2010) \emph{\mnras}, \emph{406}, 1553.

\bibitem[\protect\astroncite{\emph{{Ercolano} et~al.}}{2008}]{ercolano08}
{Ercolano} B. et~al. (2008) \emph{\apj}, \emph{688}, 398.

\bibitem[\protect\astroncite{\emph{{Ercolano} et~al.}}{2009}]{ercolano09}
{Ercolano} B. et~al. (2009) \emph{\apj}, \emph{699}, 1639.

\bibitem[\protect\astroncite{\emph{{Espaillat} et~al.}}{2012}]{espaillat12}
{Espaillat} C. et~al. (2012) \emph{\apj}, \emph{747}, 103.

\bibitem[\protect\astroncite{\emph{{Espaillat} et~al.}}{2013}]{espaillat13}
{Espaillat} C. et~al. (2013) \emph{\apj}, \emph{762}, 62.

\bibitem[\protect\astroncite{\emph{{Fang} et~al.}}{2009}]{fang09}
{Fang} M. et~al. (2009) \emph{\aap}, \emph{504}, 461.

\bibitem[\protect\astroncite{\emph{{Fedele} et~al.}}{2010}]{fedele10}
{Fedele} D. et~al. (2010) \emph{\aap}, \emph{510}, A72.

\bibitem[\protect\astroncite{\emph{{Feigelson} et~al.}}{2007}]{feigelson07}
{Feigelson} E. et~al. (2007) \emph{Protostars and Planets V}, pp. 313--328.

\bibitem[\protect\astroncite{\emph{{Feigelson} and {Decampli}}}{1981}]{fd81}
{Feigelson} E.~D. and {Decampli} W.~M. (1981) \emph{\apjl}, \emph{243}, L89.

\bibitem[\protect\astroncite{\emph{{Feigelson} and {Montmerle}}}{1999}]{fm99}
{Feigelson} E.~D. and {Montmerle} T. (1999) \emph{\araa}, \emph{37}, 363.

\bibitem[\protect\astroncite{\emph{{Font} et~al.}}{2004}]{font04}
{Font} A.~S. et~al. (2004) \emph{\apj}, \emph{607}, 890.

\bibitem[\protect\astroncite{\emph{{France} et~al.}}{2012}]{france12}
{France} K. et~al. (2012) \emph{\apj}, \emph{756}, 171.

\bibitem[\protect\astroncite{\emph{{Fromang} et~al.}}{2013}]{fromang13}
{Fromang} S. et~al. (2013) \emph{\aap}, \emph{552}, A71.

\bibitem[\protect\astroncite{\emph{{Gammie}}}{1996}]{gammie96}
{Gammie} C.~F. (1996) \emph{\apj}, \emph{457}, 355.

\bibitem[\protect\astroncite{\emph{{Geers} et~al.}}{2009}]{geers09}
{Geers} V.~C. et~al. (2009) \emph{\aap}, \emph{495}, 837.

\bibitem[\protect\astroncite{\emph{{Glassgold} et~al.}}{1997}]{glassgold97}
{Glassgold} A.~E. et~al. (1997) \emph{\apj}, \emph{480}, 344.

\bibitem[\protect\astroncite{\emph{{Glassgold} et~al.}}{2000}]{glassgold00}
{Glassgold} A.~E. et~al. (2000) \emph{Protostars and Planets IV}, p. 429.

\bibitem[\protect\astroncite{\emph{{Glassgold} et~al.}}{2007}]{glassgold07}
{Glassgold} A.~E. et~al. (2007) \emph{\apj}, \emph{656}, 515.

\bibitem[\protect\astroncite{\emph{{Gorti} and {Hollenbach}}}{2004}]{gh04}
{Gorti} U. and {Hollenbach} D. (2004) \emph{\apj}, \emph{613}, 424.

\bibitem[\protect\astroncite{\emph{{Gorti} and {Hollenbach}}}{2008}]{gh08}
{Gorti} U. and {Hollenbach} D. (2008) \emph{\apj}, \emph{683}, 287.

\bibitem[\protect\astroncite{\emph{{Gorti} and {Hollenbach}}}{2009}]{gh09}
{Gorti} U. and {Hollenbach} D. (2009) \emph{\apj}, \emph{690}, 1539.

\bibitem[\protect\astroncite{\emph{{Gorti} et~al.}}{2009}]{gorti09}
{Gorti} U. et~al. (2009) \emph{\apj}, \emph{705}, 1237.

\bibitem[\protect\astroncite{\emph{{Gorti} et~al.}}{2011}]{gorti11}
{Gorti} U. et~al. (2011) \emph{\apj}, \emph{735}, 90.

\bibitem[\protect\astroncite{\emph{{G{\"u}del} and {Naz{\'e}}}}{2009}]{gn09}
{G{\"u}del} M. and {Naz{\'e}} Y. (2009) \emph{\aapr}, \emph{17}, 309.

\bibitem[\protect\astroncite{\emph{{G{\"u}del} et~al.}}{2010}]{gudel10}
{G{\"u}del} M. et~al. (2010) \emph{\aap}, \emph{519}, A113.

\bibitem[\protect\astroncite{\emph{{Guillot} and {Hueso}}}{2006}]{guillot06}
{Guillot} T. and {Hueso} R. (2006) \emph{\mnras}, \emph{367}, L47.

\bibitem[\protect\astroncite{\emph{{Gullbring} et~al.}}{1998}]{gullbring98}
{Gullbring} E. et~al. (1998) \emph{\apj}, \emph{492}, 323.

\bibitem[\protect\astroncite{\emph{{Haisch} et~al.}}{2001}]{haisch01}
{Haisch} Jr. K.~E. et~al. (2001) \emph{\apjl}, \emph{553}, L153.

\bibitem[\protect\astroncite{\emph{{Harris} et~al.}}{2012}]{harris12}
{Harris} R.~J. et~al. (2012) \emph{\apj}, \emph{751}, 115.

\bibitem[\protect\astroncite{\emph{{Hartigan} et~al.}}{1990}]{hartigan90}
{Hartigan} P. et~al. (1990) \emph{\apjl}, \emph{354}, L25.

\bibitem[\protect\astroncite{\emph{{Hartigan} et~al.}}{1995}]{heg95}
{Hartigan} P. et~al. (1995) \emph{\apj}, \emph{452}, 736.

\bibitem[\protect\astroncite{\emph{{Hartmann} et~al.}}{1998}]{hartmann98}
{Hartmann} L. et~al. (1998) \emph{\apj}, \emph{495}, 385.

\bibitem[\protect\astroncite{\emph{{Hasegawa} and {Pudritz}}}{2012}]{hp12}
{Hasegawa} Y. and {Pudritz} R.~E. (2012) \emph{\apj}, \emph{760}, 117.

\bibitem[\protect\astroncite{\emph{{Henney} and {O'Dell}}}{1999}]{hod99}
{Henney} W.~J. and {O'Dell} C.~R. (1999) \emph{\aj}, \emph{118}, 2350.

\bibitem[\protect\astroncite{\emph{{Herczeg}}}{2007}]{herczeg07b}
{Herczeg} G.~J. (2007) in: \emph{IAU Symposium}, vol. 243, (edited by
  J.~{Bouvier} and I.~{Appenzeller}), pp. 147--154.

\bibitem[\protect\astroncite{\emph{{Herczeg} et~al.}}{2002}]{herczeg02}
{Herczeg} G.~J. et~al. (2002) \emph{\apj}, \emph{572}, 310.

\bibitem[\protect\astroncite{\emph{{Herczeg} et~al.}}{2004}]{herczeg04}
{Herczeg} G.~J. et~al. (2004) \emph{\apj}, \emph{607}, 369.

\bibitem[\protect\astroncite{\emph{{Herczeg} et~al.}}{2007}]{herczeg07}
{Herczeg} G.~J. et~al. (2007) \emph{\apj}, \emph{670}, 509.

\bibitem[\protect\astroncite{\emph{{Hern{\'a}ndez} et~al.}}{2007}]{hernandez07}
{Hern{\'a}ndez} J. et~al. (2007) \emph{\apj}, \emph{671}, 1784.

\bibitem[\protect\astroncite{\emph{{Hollenbach} and {Gorti}}}{2009}]{hg09}
{Hollenbach} D. and {Gorti} U. (2009) \emph{\apj}, \emph{703}, 1203.

\bibitem[\protect\astroncite{\emph{{Hollenbach} et~al.}}{1994}]{hollenbach94}
{Hollenbach} D. et~al. (1994) \emph{\apj}, \emph{428}, 654.

\bibitem[\protect\astroncite{\emph{{Hollenbach} et~al.}}{2005}]{hollenbach05}
{Hollenbach} D. et~al. (2005) \emph{\apj}, \emph{631}, 1180.

\bibitem[\protect\astroncite{\emph{{Hollenbach}
  et~al.}}{2000}]{hollenbach_ppiv}
{Hollenbach} D.~J. et~al. (2000) \emph{Protostars and Planets IV}, p. 401.

\bibitem[\protect\astroncite{\emph{{Hughes} and {Armitage}}}{2012}]{hughes12}
{Hughes} A.~L.~H. and {Armitage} P.~J. (2012) \emph{\mnras}, \emph{423}, 389.

\bibitem[\protect\astroncite{\emph{{Hughes} et~al.}}{2007}]{hughes07}
{Hughes} A.~M. et~al. (2007) \emph{\apj}, \emph{664}, 536.

\bibitem[\protect\astroncite{\emph{{Ingleby} et~al.}}{2009}]{ingleby09}
{Ingleby} L. et~al. (2009) \emph{\apjl}, \emph{703}, L137.

\bibitem[\protect\astroncite{\emph{{Ingleby}
  et~al.}}{2011{\natexlab{a}}}]{ingleby11}
{Ingleby} L. et~al. (2011{\natexlab{a}}) \emph{\aj}, \emph{141}, 127.

\bibitem[\protect\astroncite{\emph{{Ingleby}
  et~al.}}{2011{\natexlab{b}}}]{ingleby11a}
{Ingleby} L. et~al. (2011{\natexlab{b}}) \emph{\apj}, \emph{743}, 105.

\bibitem[\protect\astroncite{\emph{{Ingleby} et~al.}}{2012}]{ingleby12}
{Ingleby} L. et~al. (2012) \emph{\apjl}, \emph{752}, L20.

\bibitem[\protect\astroncite{\emph{{Ireland} and {Kraus}}}{2008}]{ik08}
{Ireland} M.~J. and {Kraus} A.~L. (2008) \emph{\apjl}, \emph{678}, L59.

\bibitem[\protect\astroncite{\emph{{Ivanov} et~al.}}{1999}]{ivanov99}
{Ivanov} P.~B. et~al. (1999) \emph{\mnras}, \emph{307}, 79.

\bibitem[\protect\astroncite{\emph{{Johnstone} et~al.}}{1998}]{johnstone98}
{Johnstone} D. et~al. (1998) \emph{\apj}, \emph{499}, 758.

\bibitem[\protect\astroncite{\emph{{Kamp} and {Sammar}}}{2004}]{ks04}
{Kamp} I. and {Sammar} F. (2004) \emph{\aap}, \emph{427}, 561.

\bibitem[\protect\astroncite{\emph{{Kamp} et~al.}}{2010}]{kamp10}
{Kamp} I. et~al. (2010) \emph{\aap}, \emph{510}, A18.

\bibitem[\protect\astroncite{\emph{{Kamp} et~al.}}{2011}]{kamp11}
{Kamp} I. et~al. (2011) \emph{\aap}, \emph{532}, A85.

\bibitem[\protect\astroncite{\emph{{Kastner} et~al.}}{2002}]{kastner02}
{Kastner} J.~H. et~al. (2002) \emph{\apj}, \emph{567}, 434.

\bibitem[\protect\astroncite{\emph{{Kenyon} and {Hartmann}}}{1995}]{kh95}
{Kenyon} S.~J. and {Hartmann} L. (1995) \emph{\apjs}, \emph{101}, 117.

\bibitem[\protect\astroncite{\emph{{Kim} et~al.}}{2013}]{kim13}
{Kim} K.~H. et~al. (2013) \emph{\apj}, \emph{769}, 149.

\bibitem[\protect\astroncite{\emph{{King} et~al.}}{2007}]{king07}
{King} A.~R. et~al. (2007) \emph{\mnras}, \emph{376}, 1740.

\bibitem[\protect\astroncite{\emph{{Kley} and {Nelson}}}{2012}]{kley12}
{Kley} W. and {Nelson} R.~P. (2012) \emph{\araa}, \emph{50}, 211.

\bibitem[\protect\astroncite{\emph{{Koepferl} et~al.}}{2013}]{koepferl13}
{Koepferl} C.~M. et~al. (2013) \emph{\mnras}, \emph{428}, 3327.

\bibitem[\protect\astroncite{\emph{{Kominami} and {Ida}}}{2002}]{ki02}
{Kominami} J. and {Ida} S. (2002) \emph{\icarus}, \emph{157}, 43.

\bibitem[\protect\astroncite{\emph{{K{\"o}nigl} and {Salmeron}}}{2011}]{ks11}
{K{\"o}nigl} A. and {Salmeron} R. (2011) in: \emph{Physical Processes in
  Circumstellar Disks around Young Stars}, (edited by P.~J.~V. {Garcia}), pp.
  283--352.

\bibitem[\protect\astroncite{\emph{{Kraus} and {Ireland}}}{2012}]{ki12}
{Kraus} A.~L. and {Ireland} M.~J. (2012) \emph{\apj}, \emph{745}, 5.

\bibitem[\protect\astroncite{\emph{{Kraus} et~al.}}{2008}]{kraus08}
{Kraus} A.~L. et~al. (2008) \emph{\apj}, \emph{679}, 762.

\bibitem[\protect\astroncite{\emph{{Kraus} et~al.}}{2011}]{kraus11}
{Kraus} A.~L. et~al. (2011) \emph{\apj}, \emph{731}, 8.

\bibitem[\protect\astroncite{\emph{{Kraus} et~al.}}{2012}]{kraus12}
{Kraus} A.~L. et~al. (2012) \emph{\apj}, \emph{745}, 19.

\bibitem[\protect\astroncite{\emph{{Krauss} et~al.}}{2007}]{krauss07}
{Krauss} O. et~al. (2007) \emph{\aap}, \emph{462}, 977.

\bibitem[\protect\astroncite{\emph{{Kruger} et~al.}}{2013}]{kruger13}
{Kruger} A.~J. et~al. (2013) \emph{\apj}, \emph{764}, 127.

\bibitem[\protect\astroncite{\emph{{Krumholz} et~al.}}{2013}]{krumholz13}
{Krumholz} M.~R. et~al. (2013) \emph{\apjl}, \emph{767}, L11.

\bibitem[\protect\astroncite{\emph{{Lada}}}{1987}]{lada87}
{Lada} C.~J. (1987) in: \emph{IAU Symposium}, vol. 115, (edited by
  M.~{Peimbert} and J.~{Jugaku}), pp. 1--17.

\bibitem[\protect\astroncite{\emph{{Lada} et~al.}}{2006}]{lada06}
{Lada} C.~J. et~al. (2006) \emph{\aj}, \emph{131}, 1574.

\bibitem[\protect\astroncite{\emph{{Lahuis} et~al.}}{2007}]{lahuis07}
{Lahuis} F. et~al. (2007) \emph{\apj}, \emph{665}, 492.

\bibitem[\protect\astroncite{\emph{{Lebreton} et~al.}}{2012}]{lebreton12}
{Lebreton} J. et~al. (2012) \emph{\aap}, \emph{539}, A17.

\bibitem[\protect\astroncite{\emph{{Lecavelier des Etangs}
  et~al.}}{2001}]{lecavelier01}
{Lecavelier des Etangs} A. et~al. (2001) \emph{\nat}, \emph{412}, 706.

\bibitem[\protect\astroncite{\emph{{Lega} et~al.}}{2013}]{lega13}
{Lega} E. et~al. (2013) \emph{\mnras}, \emph{431}, 3494.

\bibitem[\protect\astroncite{\emph{{Lesur} et~al.}}{2013}]{lesur13}
{Lesur} G. et~al. (2013) \emph{\aap}, \emph{550}, A61.

\bibitem[\protect\astroncite{\emph{{Li} et~al.}}{2011}]{li11}
{Li} Z.-Y. et~al. (2011) \emph{\apj}, \emph{738}, 180.

\bibitem[\protect\astroncite{\emph{{Liffman}}}{2003}]{liffman03}
{Liffman} K. (2003) \emph{\pasa}, \emph{20}, 337.

\bibitem[\protect\astroncite{\emph{{Lubow} and {D'Angelo}}}{2006}]{ld06}
{Lubow} S.~H. and {D'Angelo} G. (2006) \emph{\apj}, \emph{641}, 526.

\bibitem[\protect\astroncite{\emph{{Lugo} et~al.}}{2004}]{lugo04}
{Lugo} J. et~al. (2004) \emph{\apj}, \emph{614}, 807.

\bibitem[\protect\astroncite{\emph{{Luhman}}}{2004}]{luhman04}
{Luhman} K.~L. (2004) \emph{\apj}, \emph{617}, 1216.

\bibitem[\protect\astroncite{\emph{{Luhman} et~al.}}{2010}]{luhman10}
{Luhman} K.~L. et~al. (2010) \emph{\apjs}, \emph{189}, 353.

\bibitem[\protect\astroncite{\emph{{Lynden-Bell} and {Pringle}}}{1974}]{lbp74}
{Lynden-Bell} D. and {Pringle} J.~E. (1974) \emph{\mnras}, \emph{168}, 603.

\bibitem[\protect\astroncite{\emph{{Mamajek}}}{2009}]{mamajek09}
{Mamajek} E.~E. (2009) in: \emph{American Institute of Physics Conference
  Series}, vol. 1158, (edited by T.~{Usuda}, M.~{Tamura}, and M.~{Ishii}), pp.
  3--10.

\bibitem[\protect\astroncite{\emph{{Mann} and {Williams}}}{2010}]{mann10}
{Mann} R.~K. and {Williams} J.~P. (2010) \emph{\apj}, \emph{725}, 430.

\bibitem[\protect\astroncite{\emph{{Martin} et~al.}}{2007}]{martin07}
{Martin} R.~G. et~al. (2007) \emph{\mnras}, \emph{378}, 1589.

\bibitem[\protect\astroncite{\emph{{Marzari} et~al.}}{2010}]{marzari10}
{Marzari} F. et~al. (2010) \emph{\aap}, \emph{514}, L4.

\bibitem[\protect\astroncite{\emph{{Masset} et~al.}}{2006}]{masset06}
{Masset} F.~S. et~al. (2006) \emph{\apj}, \emph{642}, 478.

\bibitem[\protect\astroncite{\emph{{Mathews} et~al.}}{2010}]{mathews10}
{Mathews} G.~S. et~al. (2010) \emph{\aap}, \emph{518}, L127.

\bibitem[\protect\astroncite{\emph{{Mathews} et~al.}}{2012}]{mathews12}
{Mathews} G.~S. et~al. (2012) \emph{\apj}, \emph{745}, 23.

\bibitem[\protect\astroncite{\emph{{Matsumura} et~al.}}{2009}]{matsumura09}
{Matsumura} S. et~al. (2009) \emph{\apj}, \emph{691}, 1764.

\bibitem[\protect\astroncite{\emph{{Matsuyama}
  et~al.}}{2003{\natexlab{a}}}]{matsuyama03b}
{Matsuyama} I. et~al. (2003{\natexlab{a}}) \emph{\apjl}, \emph{585}, L143.

\bibitem[\protect\astroncite{\emph{{Matsuyama}
  et~al.}}{2003{\natexlab{b}}}]{matsuyama03a}
{Matsuyama} I. et~al. (2003{\natexlab{b}}) \emph{\apj}, \emph{582}, 893.

\bibitem[\protect\astroncite{\emph{{Matsuyama} et~al.}}{2009}]{matsuyama09}
{Matsuyama} I. et~al. (2009) \emph{\apj}, \emph{700}, 10.

\bibitem[\protect\astroncite{\emph{{Mayor} et~al.}}{2013}]{mayor13}
{Mayor} M. et~al. (2013) \emph{\aap}, \emph{{submitted}} (arXiv:1109.2497).

\bibitem[\protect\astroncite{\emph{{McCaughrean} and {O'Dell}}}{1996}]{mod96}
{McCaughrean} M.~J. and {O'Dell} C.~R. (1996) \emph{\aj}, \emph{111}, 1977.

\bibitem[\protect\astroncite{\emph{{Mesa-Delgado}
  et~al.}}{2012}]{mesa-delgado12}
{Mesa-Delgado} A. et~al. (2012) \emph{\mnras}, \emph{426}, 614.

\bibitem[\protect\astroncite{\emph{{Moeckel} and {Armitage}}}{2012}]{moeckel12}
{Moeckel} N. and {Armitage} P.~J. (2012) \emph{\mnras}, \emph{419}, 366.

\bibitem[\protect\astroncite{\emph{{Mohanty} et~al.}}{2013}]{mohanty13}
{Mohanty} S. et~al. (2013) \emph{\apj}, \emph{773}, 168.

\bibitem[\protect\astroncite{\emph{{Mordasini} et~al.}}{2012}]{mordasini12}
{Mordasini} C. et~al. (2012) \emph{\aap}, \emph{547}, A112.

\bibitem[\protect\astroncite{\emph{{Morishima}}}{2012}]{morishima12}
{Morishima} R. (2012) \emph{\mnras}, \emph{420}, 2851.

\bibitem[\protect\astroncite{\emph{{Muzerolle} et~al.}}{1998}]{muzerolle98}
{Muzerolle} J. et~al. (1998) \emph{\aj}, \emph{116}, 2965.

\bibitem[\protect\astroncite{\emph{{Muzerolle} et~al.}}{2000}]{muzerolle00}
{Muzerolle} J. et~al. (2000) \emph{\apjl}, \emph{535}, L47.

\bibitem[\protect\astroncite{\emph{{Muzerolle} et~al.}}{2001}]{muzerolle01}
{Muzerolle} J. et~al. (2001) \emph{\apj}, \emph{550}, 944.

\bibitem[\protect\astroncite{\emph{{Muzerolle} et~al.}}{2005}]{muzerolle05}
{Muzerolle} J. et~al. (2005) \emph{\apj}, \emph{625}, 906.

\bibitem[\protect\astroncite{\emph{{Nagasawa} et~al.}}{2005}]{nagasaw05}
{Nagasawa} M. et~al. (2005) \emph{\apj}, \emph{635}, 578.

\bibitem[\protect\astroncite{\emph{{Najita} and {Williams}}}{2005}]{najita05}
{Najita} J. and {Williams} J.~P. (2005) \emph{\apj}, \emph{635}, 625.

\bibitem[\protect\astroncite{\emph{{Najita}
  et~al.}}{2007{\natexlab{a}}}]{najita07}
{Najita} J.~R. et~al. (2007{\natexlab{a}}) \emph{\mnras}, \emph{378}, 369.

\bibitem[\protect\astroncite{\emph{{Najita}
  et~al.}}{2007{\natexlab{b}}}]{najita_ppv}
{Najita} J.~R. et~al. (2007{\natexlab{b}}) \emph{Protostars and Planets V}, pp.
  507--522.

\bibitem[\protect\astroncite{\emph{{Najita} et~al.}}{2009}]{najita09}
{Najita} J.~R. et~al. (2009) \emph{\apj}, \emph{697}, 957.

\bibitem[\protect\astroncite{\emph{{O'Dell} et~al.}}{1993}]{odell93}
{O'Dell} C.~R. et~al. (1993) \emph{\apj}, \emph{410}, 696.

\bibitem[\protect\astroncite{\emph{{Osterbrock} and
  {Ferland}}}{2006}]{osterbrock_agn2}
{Osterbrock} D.~E. and {Ferland} G.~J. (2006) \emph{{Astrophysics of gaseous
  nebulae and active galactic nuclei}}, University Science Books; USA.

\bibitem[\protect\astroncite{\emph{{Owen} and {Clarke}}}{2012}]{oc12}
{Owen} J.~E. and {Clarke} C.~J. (2012) \emph{\mnras}, \emph{426}, L96.

\bibitem[\protect\astroncite{\emph{{Owen} et~al.}}{2010}]{owen10}
{Owen} J.~E. et~al. (2010) \emph{\mnras}, \emph{401}, 1415.

\bibitem[\protect\astroncite{\emph{{Owen} et~al.}}{2011}]{owen11}
{Owen} J.~E. et~al. (2011) \emph{\mnras}, \emph{412}, 13.

\bibitem[\protect\astroncite{\emph{{Owen} et~al.}}{2012}]{owen12}
{Owen} J.~E. et~al. (2012) \emph{\mnras}, \emph{422}, 1880.

\bibitem[\protect\astroncite{\emph{{Owen} et~al.}}{2013{\natexlab{a}}}]{owen13}
{Owen} J.~E. et~al. (2013{\natexlab{a}}) \emph{\mnras}, \emph{434}, 3378.

\bibitem[\protect\astroncite{\emph{{Owen}
  et~al.}}{2013{\natexlab{b}}}]{owen13b}
{Owen} J.~E. et~al. (2013{\natexlab{b}}) \emph{\mnras}, \emph{{in press}} (arXiv:1309.0508).

\bibitem[\protect\astroncite{\emph{{Owen} et~al.}}{1999}]{owen99}
{Owen} T. et~al. (1999) \emph{\nat}, \emph{402}, 269.

\bibitem[\protect\astroncite{\emph{{Padgett} et~al.}}{2006}]{padgett06}
{Padgett} D.~L. et~al. (2006) \emph{\apj}, \emph{645}, 1283.

\bibitem[\protect\astroncite{\emph{{Pascucci} and {Sterzik}}}{2009}]{ps09}
{Pascucci} I. and {Sterzik} M. (2009) \emph{\apj}, \emph{702}, 724.

\bibitem[\protect\astroncite{\emph{{Pascucci} and {Tachibana}}}{2010}]{pt10}
{Pascucci} I. and {Tachibana} S. (2010) in: \emph{Protoplanetary Dust:
  Astrophysical and Cosmochemical Perspectives}, (edited by D.~A. {Apai} and
  D.~S. {Lauretta}), pp. 263--298.

\bibitem[\protect\astroncite{\emph{{Pascucci} et~al.}}{2006}]{pascucci06}
{Pascucci} I. et~al. (2006) \emph{\apj}, \emph{651}, 1177.

\bibitem[\protect\astroncite{\emph{{Pascucci} et~al.}}{2007}]{pascucci07}
{Pascucci} I. et~al. (2007) \emph{\apj}, \emph{663}, 383.

\bibitem[\protect\astroncite{\emph{{Pascucci} et~al.}}{2011}]{pascucci11}
{Pascucci} I. et~al. (2011) \emph{\apj}, \emph{736}, 13.

\bibitem[\protect\astroncite{\emph{{Pascucci} et~al.}}{2012}]{pascucci12}
{Pascucci} I. et~al. (2012) \emph{\apjl}, \emph{751}, L42.

\bibitem[\protect\astroncite{\emph{{Pinilla} et~al.}}{2012}]{pinilla12}
{Pinilla} P. et~al. (2012) \emph{\aap}, \emph{538}, A114.

\bibitem[\protect\astroncite{\emph{{Pollack} et~al.}}{1996}]{pollack96}
{Pollack} J.~B. et~al. (1996) \emph{\icarus}, \emph{124}, 62.

\bibitem[\protect\astroncite{\emph{{Pontoppidan} et~al.}}{2011}]{pontoppidan11}
{Pontoppidan} K.~M. et~al. (2011) \emph{\apj}, \emph{733}, 84.

\bibitem[\protect\astroncite{\emph{{Pott} et~al.}}{2010}]{pott10}
{Pott} J.-U. et~al. (2010) \emph{\apj}, \emph{710}, 265.

\bibitem[\protect\astroncite{\emph{{Quillen} et~al.}}{2004}]{quillen04}
{Quillen} A.~C. et~al. (2004) \emph{\apjl}, \emph{612}, L137.

\bibitem[\protect\astroncite{\emph{{Ratzka} et~al.}}{2007}]{ratzka07}
{Ratzka} T. et~al. (2007) \emph{\aap}, \emph{471}, 173.

\bibitem[\protect\astroncite{\emph{{Rice} et~al.}}{2003}]{rice03}
{Rice} W.~K.~M. et~al. (2003) \emph{\mnras}, \emph{342}, 79.

\bibitem[\protect\astroncite{\emph{{Richling} and {Yorke}}}{1997}]{ry97}
{Richling} S. and {Yorke} H.~W. (1997) \emph{\aap}, \emph{327}, 317.

\bibitem[\protect\astroncite{\emph{{Richling} and {Yorke}}}{2000}]{ry00}
{Richling} S. and {Yorke} H.~W. (2000) \emph{\apj}, \emph{539}, 258.

\bibitem[\protect\astroncite{\emph{{Rigliaco} et~al.}}{2013}]{rigliaco13}
{Rigliaco} E. et~al. (2013) \emph{\apj}, \emph{772}, 60.

\bibitem[\protect\astroncite{\emph{{Roberge} et~al.}}{2005}]{roberge05}
{Roberge} A. et~al. (2005) \emph{\apjl}, \emph{626}, L105.

\bibitem[\protect\astroncite{\emph{{Rosenfeld} et~al.}}{2012}]{rosenfeld12}
{Rosenfeld} K.~A. et~al. (2012) \emph{\apj}, \emph{759}, 119.

\bibitem[\protect\astroncite{\emph{{Rosotti} et~al.}}{2013}]{rosotti13}
{Rosotti} G.~P. et~al. (2013) \emph{\mnras}, \emph{430}, 1392.

\bibitem[\protect\astroncite{\emph{{Sacco} et~al.}}{2012}]{sacco12}
{Sacco} G.~G. et~al. (2012) \emph{\apj}, \emph{747}, 142.

\bibitem[\protect\astroncite{\emph{{Salmeron} et~al.}}{2011}]{salmeron11}
{Salmeron} R. et~al. (2011) \emph{\mnras}, \emph{412}, 1162.

\bibitem[\protect\astroncite{\emph{{Sargent} and {Beckwith}}}{1987}]{sb87}
{Sargent} A.~I. and {Beckwith} S. (1987) \emph{\apj}, \emph{323}, 294.

\bibitem[\protect\astroncite{\emph{{Scally} and {Clarke}}}{2001}]{sc01}
{Scally} A. and {Clarke} C. (2001) \emph{\mnras}, \emph{325}, 449.

\bibitem[\protect\astroncite{\emph{{Schindhelm} et~al.}}{2012}]{schindhelm12}
{Schindhelm} E. et~al. (2012) \emph{\apjl}, \emph{756}, L23.

\bibitem[\protect\astroncite{\emph{{Sekiya}}}{1998}]{sekiya98}
{Sekiya} M. (1998) \emph{\icarus}, \emph{133}, 298.

\bibitem[\protect\astroncite{\emph{{Shakura} and {Sunyaev}}}{1973}]{ss73}
{Shakura} N.~I. and {Sunyaev} R.~A. (1973) \emph{\aap}, \emph{24}, 337.

\bibitem[\protect\astroncite{\emph{{Shu} et~al.}}{1993}]{shu93}
{Shu} F.~H. et~al. (1993) \emph{\icarus}, \emph{106}, 92.

\bibitem[\protect\astroncite{\emph{{Shu} et~al.}}{2007}]{shu07}
{Shu} F.~H. et~al. (2007) \emph{\apj}, \emph{665}, 535.

\bibitem[\protect\astroncite{\emph{{Sicilia-Aguilar} et~al.}}{2010}]{sicilia10}
{Sicilia-Aguilar} A. et~al. (2010) \emph{\apj}, \emph{710}, 597.

\bibitem[\protect\astroncite{\emph{{Simon} et~al.}}{2013}]{simon13}
{Simon} J.~B. et~al. (2013) \emph{\apj}, \emph{775}, 73.

\bibitem[\protect\astroncite{\emph{{Simon} and {Prato}}}{1995}]{sp95}
{Simon} M. and {Prato} L. (1995) \emph{\apj}, \emph{450}, 824.

\bibitem[\protect\astroncite{\emph{{Simon} et~al.}}{2000}]{simon00}
{Simon} M. et~al. (2000) \emph{\apj}, \emph{545}, 1034.

\bibitem[\protect\astroncite{\emph{{Stelzer} et~al.}}{2013}]{stelzer13}
{Stelzer} B. et~al. (2013) \emph{\mnras}, \emph{431}, 2063.

\bibitem[\protect\astroncite{\emph{{St{\"o}rzer} and
  {Hollenbach}}}{1999}]{sh99}
{St{\"o}rzer} H. and {Hollenbach} D. (1999) \emph{\apj}, \emph{515}, 669.

\bibitem[\protect\astroncite{\emph{{Strom} et~al.}}{1989}]{strom89}
{Strom} K.~M. et~al. (1989) \emph{\aj}, \emph{97}, 1451.

\bibitem[\protect\astroncite{\emph{{Suzuki} and {Inutsuka}}}{2009}]{suzuki09}
{Suzuki} T.~K. and {Inutsuka} S.-i. (2009) \emph{\apjl}, \emph{691}, L49.

\bibitem[\protect\astroncite{\emph{{Suzuki} et~al.}}{2010}]{suzuki10}
{Suzuki} T.~K. et~al. (2010) \emph{\apj}, \emph{718}, 1289.

\bibitem[\protect\astroncite{\emph{{Szul{\'a}gyi} et~al.}}{2012}]{szulagyi12}
{Szul{\'a}gyi} J. et~al. (2012) \emph{\apj}, \emph{759}, 47.

\bibitem[\protect\astroncite{\emph{{Takeuchi} and {Lin}}}{2005}]{tl05}
{Takeuchi} T. and {Lin} D.~N.~C. (2005) \emph{\apj}, \emph{623}, 482.

\bibitem[\protect\astroncite{\emph{{Takeuchi} et~al.}}{2005}]{takeuchi05}
{Takeuchi} T. et~al. (2005) \emph{\apj}, \emph{627}, 286.

\bibitem[\protect\astroncite{\emph{{Thi} et~al.}}{2010}]{thi10}
{Thi} W.-F. et~al. (2010) \emph{\aap}, \emph{518}, L125.

\bibitem[\protect\astroncite{\emph{{Throop} and {Bally}}}{2005}]{throop05}
{Throop} H.~B. and {Bally} J. (2005) \emph{\apjl}, \emph{623}, L149.

\bibitem[\protect\astroncite{\emph{{Tielens} and {Hollenbach}}}{1985}]{th85}
{Tielens} A.~G.~G.~M. and {Hollenbach} D. (1985) \emph{\apj}, \emph{291}, 722.

\bibitem[\protect\astroncite{\emph{{Torres} et~al.}}{2008}]{torres08}
{Torres} C.~A.~O. et~al. (2008) in: \emph{Handbook of Star Forming Regions,
  Volume II}, (edited by B.~{Reipurth}), p. 757.

\bibitem[\protect\astroncite{\emph{{van Boekel} et~al.}}{2009}]{vanboekel09}
{van Boekel} R. et~al. (2009) \emph{\aap}, \emph{497}, 137.

\bibitem[\protect\astroncite{\emph{{van der Marel}
  et~al.}}{2013}]{vandermarel13}
{van der Marel} N. et~al. (2013) \emph{Science}, \emph{340}, 1199.

\bibitem[\protect\astroncite{\emph{{van Leeuwen}}}{2007}]{vanleeuwen07}
{van Leeuwen} F. (2007) \emph{\aap}, \emph{474}, 653.

\bibitem[\protect\astroncite{\emph{{Veras} and {Armitage}}}{2004}]{veras04}
{Veras} D. and {Armitage} P.~J. (2004) \emph{\mnras}, \emph{347}, 613.

\bibitem[\protect\astroncite{\emph{{Wahhaj} et~al.}}{2010}]{wahhaj10}
{Wahhaj} Z. et~al. (2010) \emph{\apj}, \emph{724}, 835.

\bibitem[\protect\astroncite{\emph{{Welsh} et~al.}}{2012}]{welsh12}
{Welsh} W.~F. et~al. (2012) \emph{\nat}, \emph{481}, 475.

\bibitem[\protect\astroncite{\emph{{Williams} and {Cieza}}}{2011}]{wc11}
{Williams} J.~P. and {Cieza} L.~A. (2011) \emph{\araa}, \emph{49}, 67.

\bibitem[\protect\astroncite{\emph{{Wilner} et~al.}}{2005}]{wilner05}
{Wilner} D.~J. et~al. (2005) \emph{\apjl}, \emph{626}, L109.

\bibitem[\protect\astroncite{\emph{{Woitke} et~al.}}{2009}]{woitke09}
{Woitke} P. et~al. (2009) \emph{\aap}, \emph{501}, 383.

\bibitem[\protect\astroncite{\emph{{Woitke} et~al.}}{2010}]{woitke10}
{Woitke} P. et~al. (2010) \emph{\mnras}, \emph{405}, L26.

\bibitem[\protect\astroncite{\emph{{Woitke} et~al.}}{2011}]{woitke11}
{Woitke} P. et~al. (2011) \emph{\aap}, \emph{534}, A44.

\bibitem[\protect\astroncite{\emph{{Wolk} and {Walter}}}{1996}]{ww96}
{Wolk} S.~J. and {Walter} F.~M. (1996) \emph{\aj}, \emph{111}, 2066.

\bibitem[\protect\astroncite{\emph{{Wright} et~al.}}{2011}]{wright11}
{Wright} J.~T. et~al. (2011) \emph{\pasp}, \emph{123}, 412.

\bibitem[\protect\astroncite{\emph{{Wyatt}}}{2008}]{wyatt08}
{Wyatt} M.~C. (2008) \emph{\araa}, \emph{46}, 339.

\bibitem[\protect\astroncite{\emph{{Yang} et~al.}}{2012}]{yang12}
{Yang} H. et~al. (2012) \emph{\apj}, \emph{744}, 121.

\bibitem[\protect\astroncite{\emph{{Youdin} and {Shu}}}{2002}]{youdin02}
{Youdin} A.~N. and {Shu} F.~H. (2002) \emph{\apj}, \emph{580}, 494.

\bibitem[\protect\astroncite{\emph{{Zhu} et~al.}}{2012}]{zhu12}
{Zhu} Z. et~al. (2012) \emph{\apj}, \emph{755}, 6.

\bibitem[\protect\astroncite{\emph{{Zuckerman} et~al.}}{1995}]{zuckerman95}
{Zuckerman} B. et~al. (1995) \emph{\nat}, \emph{373}, 494.

\end{thebibliography}

\end{document}